%% file: heavyAxion_v9.tex
\documentclass[aps,prd,nofootinbib,showpacs,preprintnumbers,longbibliography,amssymb,11pt]{revtex4-1} 
\usepackage[english]{babel}
\usepackage[utf8]{inputenc}
\usepackage[sort&compress]{natbib}
\usepackage[dvipsnames]{xcolor}
\usepackage{graphicx}
\usepackage{epsfig}
\usepackage{dcolumn}
\usepackage{bm}
\usepackage{amsmath}
\usepackage{physics}
\usepackage{slashed}
\usepackage{siunitx}
\usepackage{float}
\usepackage{hyperref}
\usepackage{caption}
\usepackage{subcaption}
\usepackage{tikz}
\usepackage[normalem]{ulem}

\usetikzlibrary{arrows}
\usetikzlibrary{decorations.text}
\usetikzlibrary{backgrounds}
\usetikzlibrary{shapes.misc}
\tikzset{cross/.style={cross out, draw=black, fill=none, minimum size=2*(#1-\pgflinewidth), inner sep=0pt, outer sep=0pt}, cross/.default={2pt}}

\usepackage{pgfplots}
\DeclareUnicodeCharacter{2212}{−}
\usepgfplotslibrary{groupplots,dateplot}
\usetikzlibrary{patterns,shapes.arrows}
\pgfplotsset{compat=newest}

\newcommand{\eqnref}[1]{Eq.~(\ref{eq:#1})}
\newcommand{\eqnsref}[2]{Eqs.~(\ref{eq:#1}) and (\ref{eq:#2})}
\newcommand{\secref}[1]{Sec.~\ref{sec:#1}}

\newcommand{\subsecref}[1]{Subsec.~\ref{subsec:#1}}

\newcommand{\appref}[1]{Appendix~\ref{sec:#1}}
\newcommand{\figref}[1]{Fig.~\ref{fig:#1}}
\newcommand{\figsref}[2]{Figs.~\ref{fig:#1} and \ref{fig:#2}}

\newcommand{\tableref}[1]{Table~\ref{table:#1}}

\begin{document}

\preprint{MITP-22-058}

\title{Supersizing axions with small size instantons}





\author{Alexey Kivel}
\email{alkivel@uni-mainz.de}
\affiliation{PRISMA+ Cluster of Excellence \& Mainz Institute for
  Theoretical Physics, Johannes Gutenberg University, 55099 Mainz,
  Germany}

\author{Julien Laux}
\email{jlaux01@uni-mainz.de}
\affiliation{PRISMA+ Cluster of Excellence \& Mainz Institute for
  Theoretical Physics, Johannes Gutenberg University, 55099 Mainz,
  Germany}

\author{Felix Yu}
\email{yu001@uni-mainz.de}
\affiliation{PRISMA+ Cluster of Excellence \& Mainz Institute for
  Theoretical Physics, Johannes Gutenberg University, 55099 Mainz,
  Germany}

\begin{abstract}
\vspace*{1cm}

We construct a new framework to calculate the enhancement of axion masses and concomitant effects on axion-meson mixing arising from small size instantons (SSIs), which originate in models featuring an extended color gauge symmetry.  The framework is based on an explicit evaluation of 't Hooft determinantal operators that partition into instanton amplitudes, affording a more precise determination of the axion-diphoton coupling than previous results.  Using an explicit model first presented in Ref.~\cite{Gaillard:2018xgk}, we demonstrate that axions solving the strong CP problem can have electroweak scale masses and higher, driven by SSI effects.  Such collider axions are prime targets for resonance searches at the Large Hadron Collider and afford a unique anchor for motivating extended color symmetries.

\end{abstract}


\maketitle

\section{Introduction}
\label{sec:Introduction}

After decades of pursuit, the Peccei-Quinn symmetry~\cite{Peccei:1977hh} and its corresponding axion particle remain the most well-studied and most viable solution to the strong CP problem of the Standard Model.  In the Standard Model, the strong CP problem stems from the undetermined relation between the possible overall phase of the quark Yukawa matrices relative to the Higgs vacuum expectation value and the unknown $\theta$ parameter characterizing the vacuum angle of quantum chromodynamics (QCD).  These phases combine into the physical observable $\overline{\theta}$, whose magnitude is constrained to be $\lesssim 10^{-10}$ from searches for an electric dipole moment of the neutron (nEDM)~\cite{Baker:2006ts, Afach:2015sja, ParticleDataGroup:2020ssz}.

Experimentally, the dominant interaction of the QCD axion with the Standard Model is expected to be its coupling to photons, leading to a breathtaking suite of experimental searches spanning decades in possible axion masses and axion-photon couplings~\cite{ParticleDataGroup:2020ssz}.  In vanilla QCD axion models, such as the historical Weinberg-Wilczek model~\cite{Weinberg:1977ma, Wilczek:1977pj} as well as the KSVZ~\cite{Kim:1979if, Shifman:1979if} and DFSZ~\cite{Dine:1981rt, Zhitnitsky:1980tq} models, the product of the axion mass $m_a$ and its decay constant $F_a$ scales as the product of the pion mass and its decay constant, connected via the topological susceptibility of QCD.  Correspondingly, these models lie on diagonal bands in the familiar $( m_a, G_{a\gamma\gamma})$ plane of QCD axion searches, where the limited spread of the bands reflects the possible model-dependence in the axion-photon coupling on fermions that realize the Peccei-Quinn symmetry at high scales and are electrically charged.

Recently, there has been a renewed discussion of ultraviolet (UV) effects on the axion mass and diphoton coupling, specifically focusing on new contributions to the axion mass from high scale instantons.  In the standard story, the axion, as the Goldstone boson of an anomalous global Peccei-Quinn symmetry, $U(1)_{\text{PQ}}$~\cite{Peccei:1977hh, Weinberg:1977ma, Wilczek:1977pj}, gains an instanton-induced potential from the topological susceptibility of QCD, and hence dynamically drives the observable $\bar{\theta}$ parameter to zero.  On the other hand, the global nature of the PQ symmetry is unprotected by possible UV operators, especially those situated at the Planck scale, leading to the axion quality problem~\cite{Barr:1992qq, Kamionkowski:1992mf, Ghigna:1992iv} and a possible spoiling of the PQ mechanism.  As a countermeasure, models have been proposed invoking compositeness~\cite{Kim:1984pt, Choi:1985cb, Randall:1992ut, Rubakov:1997vp}, extra dimensions~\cite{Choi:2003wr} or string theory~\cite{Svrcek:2006yi} to buffer the axion from UV effects and ensure a high quality axion.  For a recent overview of axion models, see Ref.~\cite{DiLuzio:2020wdo}.


In composite axion models specifically, the shallowness of the axion potential is heightened to protect the PQ mechanism, while an additional axi-eta degree of freedom from a high scale confining gauge group mixes with the nominal axion in a seesaw-type mechanism to leave the axion mass eigenvalue light.  The complication of additional Goldstone bosons that mix with the nominal axion degree of freedom is best understood from the point of view of their respective Peccei-Quinn symmetries, which are not uniquely defined.  Instead, anomalous global $U(1)$ symmetries can always be redefined via a linear combination with any gauged $U(1)$ symmetry, which can complicate the understanding of the low energy dynamics.  For example, in composite axion models, there are two anomalous global symmetries which get broken and are therefore associated with a Goldstone boson: the first is the usual PQ symmetry at the high scale $F_a$ and the second is the axial $U(1)_A$ at the low scale $F_{\eta'} \sim \Lambda_{\text{QCD}}$.  Even with this large separation of scales, we take advantage of 't Hooft anomaly matching to study the pseudo-Nambu Goldstone mass matrix at the low energy scale.  The resulting diagonalization of the mass matrix then reflects the ideal basis to study the spontaneous breaking of the corresponding PQ symmetries, since they become orthogonal charge assignments when the pseudo-Nambu Goldstone bosons (PNGBs) are diagonalized.  A critical phenomenological consequence of the mixing effects between the Goldstones is the obvious modification of the diphoton coupling of the axion mass eigenstate.

Aside from the necessary diagonalization of PNGB mass matrices in composite axion models, there have been continued developments in the study of anomalous symmetry breaking.  In Ref.~\cite{Agrawal:2017ksf, Agrawal:2017evu}, Agrawal and Howe showed that a non-trivial embedding of QCD into a new non-Abelian gauge symmetry at a high scale can result in small size instantons (SSI) that strongly influence the PQ symmetry breaking and axion potential.  The magnitude of the SSI effect on axion properties depends significantly on the details of the embedding of the color gauge group in its ultraviolet completion, as discussed in Ref.~\cite{Csaki:2019vte}, and concrete studies of enlarged QCD color groups or color unification were performed in Ref.~\cite{Gherghetta:2016fhp, Gaillard:2018xgk, Valenti:2022tsc}.

The effect on the axion mass from a non-trivial embedding of QCD into larger gauge groups has historically been a rich subject (see, e.g. Refs.~\cite{Holdom:1982ex, Randall:1992ut, Rubakov:1997vp}), where the axion mass is recalculated accounting for the effect of different UV operators.  In Ref.~\cite{Rubakov:1997vp}, for example, the UV compositeness structure also addresses the axion quality problem, as previously mentioned.  Intuitively, embedding QCD into larger gauge groups invites the question of how the QCD $\theta$ angle is inherited from the theta angles of the possible UV gauge groups.  One simple solution is a $\mathbb{Z}_2$ mirror copy of the SM, which relates the theta parameters and doubles the matter content, apart from the axion degree of freedom~\cite{Berezhiani:2000gh, Hook:2014cda, Fukuda:2015ana, Dimopoulos:2016lvn, Hook:2018dlk, Hook:2019qoh}. Consequently, the axion receives a potentially large contribution to its mass from instantons of the mirror copy of QCD. 

In much of the previous literature discussing UV instantons, the calculations have used a modified axion potential based on Ref.~\cite{Randall:1992ut} and dimensional analysis arguments.  While a modified axion potential constructed this way can readily justify a possible mass enhancement for the axion, it is insufficient to derive phenomenological properties of the axion, such as axion-hadron or axion-diphoton couplings, and dimensional analysis arguments may miss possible features such as a seesaw-like mechanism that renders an axion very light.
Also, since the axion potential from instantons can provide a mixing to other, possibly new, pseudo-scalars of the theory, similar to $\pi^0$-$\eta'$-axion mixing in QCD, it is important to include mass mixing effects from UV-instantons into the whole picture.  Clearly, since the axion-photon coupling is the main target of a majority of axion detection experiments~\cite{ParticleDataGroup:2020ssz}, it is of fundamental importance to understand the extent to which UV-instantons can alter this coupling as well as the axion mass.  

To that end, we present an updated framework of calculating the effects from UV instantons on QCD axions, which we adopt and further develop based on the approach in Refs.~\cite{Kim:2006aq, Kim:2008hd}.  Namely, we calculate the dynamical change in the expectation value $\langle G\tilde{G}\rangle$ by including the axion dependence explicitly as a phase of the instanton vertex and similarly for other pseudo-scalars of the underlying theory.  In contrast to a Wilsonian approach, our approach derives an effective axion potential that includes strong CP-violating operators from different scales simultaneously.  We therefore account for possible mass mixing effects in the UV as well as in the IR, giving us the opportunity to study the SSI sensitivity of $G_{a\gamma\gamma}$ for the first time. 

There is another technical advantage in calculating axion properties with $\langle G\tilde{G}\rangle$. In previous calculations ({\it e.g.} Refs.~\cite{Georgi:1986df, Kaplan:1985dv}), the axion is matched to an effective low-energy Lagrangian by rotating the axion field into the quark mass matrix, simultaneously rotating away the $G\tilde{G}$ term.  In models with a non-trivial ultraviolet completion of the color gauge group, however, the relation of $\theta$ of QCD to other possible $\theta'$ parameters in the UV is unclear and hence can spoil the above approach, since other $G'\tilde{G}'$ terms may remain after such a rotation.  In this new approach, we are able to solve the dilemma by making use of the 't Hooft anomaly matching to take the expectation value of all CP-violating gauge terms simultaneously. After developing the necessary steps and cross checking intermediate results, we apply our formalism to a concrete example of a color unification model~\cite{Gaillard:2018xgk}, where $SU(3)_c$ is embedded together with another non-Abelian gauge group into $SU(6)$ and where SSI effects from a $SU(3')$ gauge group affect the axion potential.

In~\secref{instantons}, we review the classic $U(1)$ problem of the $\eta'$ mass in QCD and its resolution from instanton effects.  We also include a discussion of how QCD instanton effects are distributed when an axion degree of freedom is also present.  In~\secref{generalMethod}, we review and improve a robust calculational framework (originally presented in Ref.~\cite{Kim:2008hd}) for incorporating instanton effects in the mass spectrum of pseudo-Nambu Goldstone bosons of mesons and axion degrees of framework.  The key benefit of this approach is that axion and axieta fields from high scale extensions of the color gauge symmetry are accurately mixed with the low energy QCD chiral Lagrangian meson states.  We also demonstrate that this framework reproduces results from the traditional approach based on quark mass matrix rotations~\cite{Georgi:1986df} in the required limiting case when non-QCD instanton effects are turned off.  Then, in~\secref{enhancement}, we calculate the enhancement for the axion and axieta masses from an exemplary extended color gauge model~\cite{Gaillard:2018xgk}, accounting for the leading mixing effects with QCD mesons, and we show the wide variation possible compared to vanilla QCD axion models in the $(m_a, F_{a \gamma \gamma})$ plane.  We also comment on the expected general features of SSI effects on axion properties that extend beyond the results of the exemplary model.  Our conclusions are presented in~\secref{conclusion}.  In~\appref{genMatrix}, we include a discussion about our analytic approach to diagonalize the PGNB mass matrix.

\section{A recap of instanton effects in $SU(N)$ theories and the $U(1)$ problem in QCD}
\label{sec:instantons}
In this section we review instanton effects in $SU(N)$ theories, especially in the context of the $U(1)$-problem and their contributions to masses of pseudo-Nambu Goldstone bosons, such as the $\eta'$ meson or a standard QCD axion.  Afterwards we discuss the magnitude of instanton related effects in the general context of one or several non-Abelian gauge theories.

At present, the only immediate phenomenological handle we have on instanton effects in QCD is the mass of the $\eta'$ particle stemming from the $U(1)$ problem~\cite{Weinberg:1975ui}.  In other words, the $\eta'$ particle is the only observed PNGB of an anomalous, approximate global symmetry in the SM.  The hypothesized axion particle is also a PNGB of the anomalous, global PQ symmetry, which invites the question of how QCD instanton effects are distributed between the two anomalous symmetries.  We thus require a framework that calculates instanton effects for both the $\eta'$ and axion simultaneously.  

A thought experiment illustrating the deep connection between the $\eta'$ and axion is the fact that if the massless quark solution to the Strong CP problem were realized in Nature, the $\eta'$ plays the role of the axion field.  For instance, the global $U(3) \times U(3)$ flavor symmetry in QCD is typically broken by the quark condensate to give the observed massive PNGBs and the famous $U(1)$ problem.  If, however, all quarks are massless, then the corresponding PNGBs are also massless except for the $\eta'$ boson because of its mass contribution from instantons.  More importantly, since there are no mass terms that break the anomalous $U(1)_A \in U(3) \times U(3)$, the $U(1)_A$ symmetry can be used to render $\bar{\theta}$ unphysical.  Thus, we have succesfully reproduced a PQ symmetry with $\eta'$ being the corresponding PNGB, without the need for adding an axion degree of freedom.  From this argument, we see that an effective QCD instanton potential needs to fulfill certain conditions. Namely, it has to generate a mass splitting between the axion, $\pi_0$ and $\eta'$ that fulfills the following limiting cases.  In the limit that all quarks are massless, we have two anomalous global symmetries: $U(1)_{PQ}$ and $U(1)_A$. Since all quarks are massless, we are free to redefine the two anomalous symmetries such that we have one anomaly-free symmetry. Therefore, one of the corresponding mass eigenvalues becomes zero in this case. We would then have $m_a F_a = 0$, $m_{\eta'} F_{\eta'} > 0$, and $m_{\pi_0} F_{\pi_0} = 0$. For finite and universal quark masses and neglecting instanton effects, we would instead have $m_{\eta'} F_{\eta'} = m_{\pi_0} F_{\pi_0}$, and $m_a F_a = 0$.  We see that the topological susceptibility, which characterizes the impact of the instanton contributions from QCD, aligns with the PNGB associated with $\bar{\theta}$, which is a central feature of models where the QCD theta parameter is inherited from high scale QCD embeddings.


Of course, in the Standard Model, the massless quark solution has been excluded by lattice extractions of the quark masses~\cite{Fodor:2016bgu}.  Moreover, 't Hooft solved the $U(1)$ problem~\cite{tHooft:1976rip, tHooft:1986ooh} by matching instanton degrees of freedom from $G \tilde{G}$ to an effective operator that anomalously increases the $\eta'$ mass.  Thus, using the dilute gas approximation for the instanton calculation, we can trade the $G \tilde{G}$ operator for the effective 't Hooft determinantal operator, which is schematically written as
\begin{align}
\label{eq:Ldet1}
\frac{\theta}{32\pi^2}G^a_{\mu\nu}\tilde{G}^{a,\mu\nu}\Leftrightarrow\mathcal{L}_{\text{det}} \approx K^{4 - 3 N_f} e^{-i\theta} \prod_{i=1}^{N_f} \det( \bar{q}^i_L q^i_R) + \text{h.c.} = 
2K^{4 - 3 N_f}
\left( \prod_{i=1}^{N_f} \det( \bar{q}^i_L q^i_R) \right) \cos(\theta) \ ,
\end{align}
the index $i$ runs over all quark flavors $q_i$, the final step requires $\det( \bar{q}^i_L q^i_R ) = \det( \bar{q}^i_R q^i_L )$, and $K$ is the effective instanton amplitude~\cite{tHooft:1976rip, tHooft:1986ooh, Kim:2008hd}.  The phenomenology of mesons in low energy QCD is generally calculated using chiral effective theory and hence the instanton amplitude $K$ is extracted using the $\eta'$ mass~\cite{tHooft:1976rip, tHooft:1986ooh}. 

Intuitively, the determinantal operator arises as a consequence of the fact that the background instanton configuration necessarily changes the number of left-handed modes, $n_+$,  and right-handed modes, $n_-$, according to the topological winding number $\nu$ of the instanton,
\begin{align}
\nu = n_+ - n_- \ ,
\end{align}
which itself is a manifestation of the Atiyah-Singer index theorem~\cite{Atiyah:1963, Atiyah:1968mp} and acts as a selection rule for the fermion modes in the path integral~\cite{Fujikawa:1980eg}.

There are two contributions that let us close off the bilinears in the determinantal term, i.e. Higgs-vev insertion in the Yukawa coupling or the quark condensate from QCD, which are both sources of chiral symmetry breaking. Diagrammatically this is shown in instanton diagrams that give contributions to the potential of the PNGB below breaking scale. Because of the equivalence to the $G\tilde{G}$ term, the PNGBs associated with anomalous symmetries get an instanton contribution to their mass.  This solves the $U(1)$ problem for the $\eta'$ mass.  

In the axion case, the determinantal operator generates an approximate cosine potential that dynamically relaxes the $\bar{\theta}$ parameter to 0~\cite{Peccei:1977hh, diCortona:2015ldu}.  Since the determinantal operator reflects the instanton contributions to PNGB masses, the determinantal operator contributes to PNGBs that are non-trivially charged under either the global PQ symmetry as well as the familiar $U(1)_A$ flavor symmetry from $N_F = 3$ low energy QCD, which are both anomalous with respect to color.  Hence, the QCD instanton effects lead to mass contributions to as well as mass mixing between PNGBs.  Moreover, the quark condensate explicitly breaks the chiral symmetries of all PNGBs, except the axion.  The combination of the quark condensate and instanton-induced mixing gives the standard axion lore, $m_a^2 f_a^2 = \mathcal{T}$, where $\mathcal{T}$ is the topological susceptibility of QCD and is briefly reviewed in~\appref{topSusc}, and we identify $\mathcal{T} \approx \Lambda_{\text{QCD}}^4 \approx m_{\pi^0}^2 F_{\pi^0}^2$.  For small $\bar{\theta}$, the topological susceptibility is related to the instanton amplitude $K$ via $\mathcal{T} \sim m_u m_d m_s K$~\cite{Huang:1993cf}.  Taking into account phenomenological constraints, the axion decay constant $F_a$ is required to be much larger than the electroweak scale, typically at least $\mathcal{O}(10^{12})$~GeV~\cite{ParticleDataGroup:2020ssz}, which generally results in $m_a = 5.7 \mu\text{eV} \left( \frac{10^{12}~\text{GeV}}{F_a} \right)$~\cite{diCortona:2015ldu}.

In our approach with the determinantal operator, the magnitude of the contribution is visible in the effective instanton amplitude $K$, which was introduced in~\eqnref{Ldet1}. This amplitude consists of a weighted integral over all instanton sizes $\rho$ where each size corresponds to an energy scale $\mu=1/\rho$~\cite{tHooft:1976snw, Flynn:1987rs,Bernard:1979qt},
\begin{align}
K^{4-N_f} \sim \int \frac{\dd{\rho}}{\rho^{5-N_f}} \exp \left[ \frac{-2\pi}{\alpha(\mu)} \right] \ ,
\end{align}
where $\alpha(\mu) = \frac{g^2(\mu)}{4\pi}$ and $g(\mu)$ is the running coupling.  Given that $SU(3)_c$ is asymptotically free, the smallness of $\alpha(\mu)$ at high energy scales leads to a huge exponential suppression.  The phenomenology of the standard axion is thus dominated by large-size instantons. 

Following Refs.~\cite{Holdom:1982ex, Flynn:1987rs, Randall:1992ut, Rubakov:1997vp, Csaki:1998vv, Agrawal:2017ksf, Agrawal:2017evu,  Csaki:2019vte}, for example, it is possible, however, to circumvent the exponential suppression by invoking a new non-Abelian gauge theory $SU(N)$ that confines at a high scale and contains matter content charged under $SU(N) \times SU(3)_c$.  Alternatively, we can invoke a new non-Abelian gauge theory $SU(N)'$ that non-trivially embeds $SU(3)_c$ and also confines at a high scale.

In either case, denoting the high scale $\mu' \gg \Lambda_{\text{QCD}}$, the small-size instantons with $\rho = 1 / \mu'$ give an enhanced instanton amplitude since the running coupling $\alpha'$ is $\mathcal{O}(1)$ at this scale.  This dramatic enhancement of the small-size instanton amplitude,
\begin{align}
\left( K^\prime \right)^{4-N_f} \sim \int \frac{\dd{\rho}}{\rho^{5-N_f}} \exp \left[ \frac{-2\pi}{\alpha'(1/\rho)} \right] \ ,
\end{align}
leads to an enhancement of the corresponding axion mass by several orders of magnitude since $K' \gg K$, which we will calculate in the next section.

We can now explain how recent work has used UV-instantons as model-building blocks to enlarge the class of viable QCD axion solutions.  For example, Ref.~\cite{Agrawal:2017evu} demonstrated that the massless up-quark solution becomes viable again in the context of a non-trivial embedding of $SU(3)_c$ in a $SU(3)^3$ gauge group~\cite{Agrawal:2017evu}, where the high-scale instantons give a non-perturbative contribution to the up-quark mass.  (The massless up-quark solution has also been discussed in Ref.~\cite{Bardeen:2018fej}.)  In Ref.~\cite{Gaillard:2018xgk}, the authors discuss ''color unification'' models with an $SU(6) \times SU(3)$ gauge group, where one massless fermion in the $SU(6)$ gauge group rotates away all $\theta$ parameters and $SU(3)_c$ is embedded as a subgroup of the $SU(6)$ symmetry.  This is similar to the original construction in Ref.~\cite{Agrawal:2017ksf}, where both references argue from dimensional analysis about the structure of the axion potential and the heavy axion mass.  An important fact is that such axion models do not suffer from the low-quality problem, since the heightened axion potential protects from corrections of Planck-scale UV-operators.

As mentioned in the introduction, in order to calculate the correct mixing effects from small-size instantons, we need to derive the axion potential integrating out the $\langle G \tilde{G} \rangle$ operator, which is the subject of the next section.

\section{QCD axion low-energy properties from treating instanton effects as a phase of the $\langle G \tilde{G} \rangle$ operator}
\label{sec:generalMethod}

In this section, we review a method for calculating axion properties from first principles, based on~\cite{Kim:2008hd}.  Since the axion Lagrangian enjoys a shift symmetry, the various couplings of the axion field to the matter content and the field strengths are basis-dependent.  In particular, the axion coupling to $\langle G \tilde{G} \rangle$, which is the source for the vanilla axion mass after QCD confinement, is typically traded for a coupling to the quark bilinear~\cite{Georgi:1986df}, from which the axion mass and diphoton coupling are derived using the effective QCD chiral Lagrangian.  We review this derivation in~\subsecref{comparison}.

Since we want to account for possible small-size instanton effects, we instead eschew the traditional approach in favor the framework presented in Ref.~\cite{Kim:2008hd}, where the axion and all PNGBs are treated as phases of their respective symmetries in both instanton diagrams as well as the QCD chiral condensate.  An advantage of this approach will be that the basis-dependence of the axion couplings will manifest as mixing matrices among the different PNGBs.

Given the spontaneous breaking of the global anomalous Peccei-Quinn symmetry $U(1)_{\text{PQ}}$ at a scale $F_a$, the general effective Lagrangian for the axion and SM fields takes the form~\cite{Georgi:1986df, Kim:2008hd, Bauer:2020jbp}
\begin{align}
\nonumber
\mathcal{L} &= \frac{1}{2}\partial_\mu a \partial^\mu a+ \frac{\partial_\mu a}{F_a} \left( \sum_{i=1}^{N_f} c_1^i \bar{q}_i \gamma_\mu \gamma_5 q_i\right)- \left( \sum_{i=1}^{N_f} m_i \bar{q}_L^i e^{i c_2^{i} a/F_a} q_R^i +\text{h.c.}  \right) \\[5pt]
& -\frac{a}{F_a}\left(c_3^G \frac{g_s^2}{32\pi^2} G\tilde{G} + c_3^W \frac{g^2}{32\pi^2} W\tilde{W} + c_3^B \frac{g'^2}{32\pi^2} B\tilde{B}\right) \ ,
\label{eq:LSM}
\end{align}
where $q_i$ are the SM quarks in the mass basis and $G$, $W$, $B$ the field strengths with the respective gauge couplings $g_s$, $g$, and $g'$. We define $a$ as the dynamical part of the axion field, where the tadpole cancels $\bar{\theta}$.  Following Refs.~\cite{Kim:2008hd, Bauer:2020jbp} we introduce the effective couplings $c_1^i$, $c_2^i$ $c_3^G$, $c_3^W$, and $c_3^B$ to account for different axion models.  For example, the KSVZ model~\cite{Kim:1979if, Shifman:1979if} would have $c_1^i = c_2^i = 0$ and $c_3^G \neq 0$, since the axion solely couples to heavy quarks not present in the effective description.  Note that possible off-diagonal couplings in $c_2$ can result from either axiflavon models or CKM-induced flavor violation~\cite{Calibbi:2016hwq, Ema:2016ops, MartinCamalich:2020dfe}, but we will assume $c_2$ is diagonal in the quark mass basis in this paper.

As previously mentioned, the axion shift symmetry is manifest by performing the axial $U(1)$ quark transformation $q_i \to e^{i \alpha \gamma_5 ( a / F_a)} q_i$, simultaneously with the coupling re-definitions \begin{align}
c_1^i \rightarrow c_1^i - \alpha, \quad c_2^i \rightarrow c_2^i - 2\alpha, \quad 
c_3^G \rightarrow c_3^G + 2\alpha \ ,
\end{align}
for each individual quark, $i = 1$, $\ldots$, $6$, and the change in the $c_3^G$ term corresponds to the chiral anomaly.  For now, we ignore the concomitant changes in $c_3^W$ and $c_3^B$, which will be discussed later in context of the diphoton coupling.

\subsection{Instanton diagrams as vertex phases}
\label{subsec:method}

Studying~\eqnref{LSM}, the axion mass is generated by the breaking of the axion shift symmetry by the $\langle G \tilde{G} \rangle$ operator after QCD confinement.  Hence, following Ref.~\cite{Kim:2008hd}, we match the $G \tilde{G}$ operator to the 't Hooft determinantal operator as in~\eqnref{Ldet1}, and then sum over instanton effects by decomposing the determinantal operator into explicit products of quark bilinear  contributions.  In this way, the axion and all PNGB fields are treated as phases of the instanton diagrams expressed as quark bilinear products , according to the current transformation of each bilinear.

The starting point is the axion Lagrangian in~\eqnref{LSM} just above the QCD confinement scale, where we replace the $G \tilde{G}$ operator by the 't Hooft determinantal operator~\cite{tHooft:1976rip, tHooft:1986ooh},
\begin{align}
\mathcal{L}_{\text{det}} = 
(-1)^{N_f} K^{4-3N_f}
\left( \prod\limits_{i = 1}^{N_f} 
\det(\bar{q}_L^i q_R^i ) 
\right) e^{-i c_3^G \frac{a}{F_a}  } + \text{h.c.} \ ,
\label{eq:Ldet2}
\end{align}
where $K$ is the instanton amplitude, $N_f$ is the number of quarks with PQ-charge ($N_f$ = 3 in the usual axion story) and we emphasize that we focus on the dynamical field $a$ and drop the $\bar{\theta}$ constants.  To study the instanton contributions to the mass matrix for the axion and the PNGBs from light quarks in QCD, we necessarily need to close off the determinantal operator into color singlet contributions using all available chiral symmetry breaking vacuum expectation values (vevs).

In general, the mesons $\pi^0$, $\eta$ and $\eta'$ are Goldstone bosons that result from the spontaneous chiral symmetry breaking due to the quark condensation $\langle\bar{q}q\rangle \equiv v^3$.  We define the mesons as angular field excitations around $v^3$, assigning transformation properties according to the appropriate flavor symmetry generators.  For simplicity, we discard the $\eta$-meson and its mixing with the other mesons since it is small.  We recognize that the mass mixing from the dynamical strange quark is necessary suppressed by $1/ m_s$ since it provides the leading flavor symmetry breaking from 3-flavor QCD to 2-flavor QCD, and thus we will calculate in an approximate 3-flavor QCD where the mixing from the strange quark are ignored.  Hence, we use the approximation
\begin{align}
&\bar{u}_Lu_R\approx |\langle\bar{u}_Lu_R\rangle| \exp(i(\theta_{\pi^0}+\theta_{\eta'})) = \frac{v^3}{2} \exp(i(\theta_{\pi^0}+\theta_{\eta'})) \ ,
\label{eq:expUp3f}\\
&\bar{d}_Ld_R\approx |\langle\bar{d}_Ld_R\rangle| \exp(i(-\theta_{\pi^0}+\theta_{\eta'})) = \frac{v^3}{2} \exp(i(-\theta_{\pi^0}+\theta_{\eta'})) \ ,
\label{eq:expDown3f} \\
&\bar{s_L}s_R\approx |\langle\bar{s_L}s_R\rangle| \exp(i\theta_{\eta'}) = \frac{v^3}{2} \exp(i\theta_{\eta'}) \sim \frac{v^3}{2} \ ,
\label{eq:expStr3f}
\end{align}
for the three light quarks, where $\theta_{\pi^0} = \pi^0 / F_{\pi^0}$, $\theta_{\eta'} = \eta' / F_{\eta'}$ with decay constants $F_\pi^0$ and $F_{\eta'}$.  We see explicitly that the signs above correspond to the particle content of the meson, for example $\pi^0=(\bar{u}u-\bar{d}d)/\sqrt{2}$.

We now build up each contribution to the axion and Goldstone mass matrix from~\eqnref{LSM} where $G \tilde{G}$ is replaced by~\eqnref{Ldet2}, following the method proposed in~\cite{Kim:2006aq, Kim:2008hd}.  While the determinantal expression is IR-divergent, 't Hooft has shown that this term can be evaluated with the help of chirality changing source terms~\cite{tHooft:1976rip}, known as instanton diagrams.  In these diagrams we close fermion legs around a bubble that symbolizes the instanton using either Yukawa insertions or the quark condensate as our chirality changing source.  We necessarily work in the Higgs phase of the electroweak theory, where Yukawa insertions are expressed as quark masses~\cite{Shifman:2012zz}.

Starting with the quark mass terms in~\eqnref{LSM}, we have
\begin{align}
\label{eq:quarkMassTermPot}
\mathcal{L} &\supset - m_u \bar{u}_L e^{i c_2^u \frac{a}{F_a}} u_R - m_d \bar{d}_L e^{i c_2^d \frac{a}{F_a}} d_R - m_s \bar{s}_L e^{i c_2^s \frac{a}{F_a}} s_R + \text{ h.c.} \\
&\approx - m_u v^3 \cos(\theta_{\pi^0} + \theta_{\eta'} + c_2^u \theta_a) - m_d v^3 \cos(-\theta_{\pi^0} + \theta_{\eta'} + c_2^d \theta_a) - m_s v^3\cos(\theta_{\eta'} + c_2^s \theta_a) \ ,
\end{align}
where $\theta_a \equiv a / F_a$ is the axion field written as a phase.

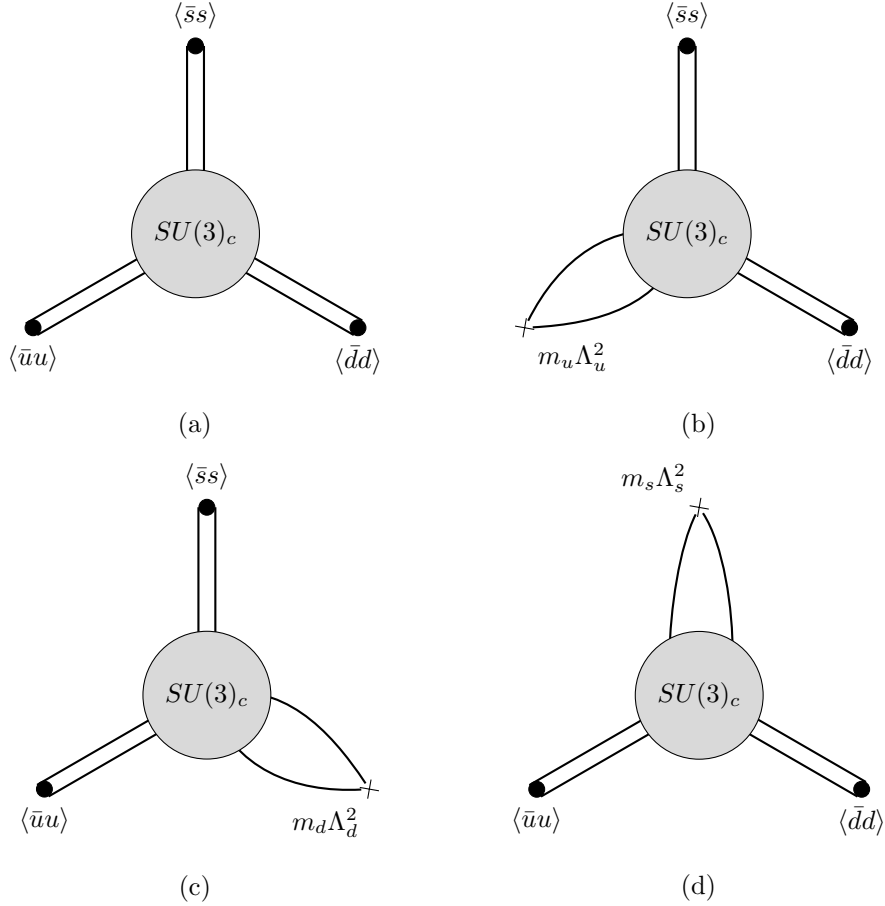
\begin{figure}[tb!]
\begin{subfigure}{0.4\textwidth}
\centering
\begin{tikzpicture}
\node[circle,draw, minimum size = 1.7cm,color = black, fill = gray!30] (center) at  (0,0) {$SU(3)_c$};
\begin{scope}[on background layer]
\node[circle,draw,inner sep =0pt, minimum size = 6pt,color = white] (helper) at  (0,0){};

\node[circle,draw,inner sep =0pt, minimum size = 6pt,color = black,fill=black, label = below:$\langle \bar{u}u\rangle$] (A) at (-2.16,-1.25){};

\node[circle,draw,inner sep =0pt, minimum size = 6pt,color = black,fill=black, label = below:$\langle \bar{d}d\rangle$] (B) at (2.16,-1.25){};

\node[circle,draw,inner sep =0pt, minimum size = 6pt,color = black,fill=black, label = above:$\langle \bar{s}s\rangle$] (C) at (0,2.5){};


\draw[thick] (helper.-60) -- (A.-60);
\draw[thick] (helper.150) -- (A.150);

\draw[thick] (helper.60) -- (B.60);
\draw[thick] (helper.240) -- (B.240);

\draw[thick] (helper.0) -- (C.0);
\draw[thick] (helper.180) -- (C.180);
\end{scope}
\end{tikzpicture}
\caption{\label{subfig:bubble1}}
\end{subfigure}
\begin{subfigure}{0.4\textwidth}
\centering
\begin{tikzpicture}
\node[circle,draw, minimum size = 1.7cm,color = black, fill = gray!30] (center) at  (0,0) {$SU(3)_c$};
\begin{scope}[on background layer]
\node[circle,draw,inner sep =0pt, minimum size = 6pt,color = white] (helper) at  (0,0){};

\node[cross=3,rotate=35,label = below:$m_u\Lambda_u^2$] (A) at (-2.16,-1.25){};

\node[circle,draw,inner sep =0pt, minimum size = 6pt,color = black,fill=black, label = below:$\langle \bar{d}d\rangle$] (B) at (2.16,-1.25){};

\node[circle,draw,inner sep =0pt, minimum size = 6pt,color = black,fill=black, label = above:$\langle \bar{s}s\rangle$] (C) at (0,2.5){};


\draw [thick] (helper) to [out=250,in=00] (A);
\draw [thick] (helper) to [out=170,in=60](A);

\draw[thick] (helper.60) -- (B.60);
\draw[thick] (helper.240) -- (B.240);

\draw[thick] (helper.0) -- (C.0);
\draw[thick] (helper.180) -- (C.180);
\end{scope}
\end{tikzpicture}
\caption{\label{subfig:bubble2}}
\end{subfigure}
\begin{subfigure}{0.4\textwidth}
{\centering
\begin{tikzpicture}
\node[circle,draw, minimum size = 1.7cm,color = black, fill = gray!30] (center) at  (0,0) {$SU(3)_c$};
\begin{scope}[on background layer]
\node[circle,draw,inner sep =0pt, minimum size = 6pt,color = white] (helper) at  (0,0){};

\node[circle,draw,inner sep =0pt, minimum size = 6pt,color = black,fill=black, label = below:$\langle \bar{u}u\rangle$] (A) at (-2.16,-1.25){};

\node[cross=3,rotate=35,label = below left:$m_d\Lambda_d^2$] (B) at (2.16,-1.25){};

\node[circle,draw,inner sep =0pt, minimum size = 6pt,color = black,fill=black, label = above:$\langle \bar{s}s\rangle$] (C) at (0,2.5){};


\draw[thick] (helper.-60) -- (A.-60);
\draw[thick] (helper.150) -- (A.150);

\draw [thick] (helper) to [out=370,in=120] (B);
\draw [thick] (helper) to [out=290,in=180](B);

\draw[thick] (helper.0) -- (C.0);
\draw[thick] (helper.180) -- (C.180);
\end{scope}
\end{tikzpicture}
\par }
\caption{\label{subfig:bubble3}}
\end{subfigure}
\begin{subfigure}{0.4\textwidth}
{\centering
\begin{tikzpicture}
\node[circle,draw, minimum size = 1.7cm,color = black, fill = gray!30] (center) at  (0,0) {$SU(3)_c$};
\begin{scope}[on background layer]
\node[circle,draw,inner sep =0pt, minimum size = 6pt,color = white] (helper) at  (0,0){};

\node[circle,draw,inner sep =0pt, minimum size = 6pt,color = black,fill=black, label = below:$\langle \bar{u}u\rangle$] (A) at (-2.16,-1.25){};

\node[circle,draw,inner sep =0pt, minimum size = 6pt,color = black,fill=black, label = below:$\langle \bar{d}d\rangle$] (B) at (2.16,-1.25){};

\node[cross=3,rotate=35,label = above:$m_s\Lambda_s^2$] (C) at (0,2.5){};

\draw[thick] (helper.-60) -- (A.-60);
\draw[thick] (helper.150) -- (A.150);

\draw[thick] (helper.60) -- (B.60);
\draw[thick] (helper.240) -- (B.240);

\draw [thick] (helper) to [out=240,in=240] (C);
\draw [thick] (helper) to [out=300,in=300](C);
\end{scope}
\end{tikzpicture}
\caption{\label{subfig:bubble4}}
\par }
\end{subfigure}
\caption[Instanton diagrams in QCD]{The leading instanton diagrams from the determinantal interaction in~\eqnref{Ldet2}, organized by their chirality-changing source terms.  Diagram~(\subref{subfig:bubble1}) uses only the quark condensate, denoted by $\bullet$, while diagrams~(\subref{subfig:bubble2})$-$(\subref{subfig:bubble4}) each use one insertion of a light quark mass, symbolized by~$\times$.
\label{fig:instaFlower}
}
\end{figure}
From~\eqnref{Ldet2}, the most relevant instanton diagrams are shown in~\figref{instaFlower}, where the various quark bilinears are closed off using the quark condensate or the mass insertion.  We neglect higher order diagrams suppressed by multiple insertions of quark masses.

We assign the instanton diagrams in~\figref{instaFlower} to separate instanton amplitudes $A_i$,
\begin{align}
\label{eq:determInteraction}
\mathcal{L}_{\text{det}} = -\frac{1}{K^5} \sum_i A_i \ , 
\end{align}
where each $A_i$ is mass dimension $9$.  Using~\eqnref{expUp3f}$-$~\eqnref{expStr3f}, the instanton diagram~\figref{instaFlower}\subref{subfig:bubble1} is calculated to be
\begin{align}
A_1 &= 
\left( \prod_i \det(\bar{q}_{i,L} \, q_{i,R}) \right) e^{-i c_3^G \theta_a} + \text{ h.c.} \ ,
\nonumber \\
& \sim \left( \frac{v^3}{2} \exp(i (\theta_{\pi^0} + \theta_{\eta'})) \right) \left( \frac{v^3}{2} \exp(i (-\theta_{\pi^0} + \theta_{\eta'})) \right) \left( \frac{v^3}{2} \right) e^{-i c_3^G \theta_a}  + \text{ h.c.} \ , \nonumber
\\[12pt]
&= \frac{v^9}{8} \left(\exp(i(2\theta_{\eta'} - c_3^G \theta_a )) + \text{h.c.} \right) = \frac{v^9}{4} \cos(2\theta_{\eta'} - c_3^G \theta_a) \ ,
\label{eq:A1instanton}
\end{align}
following our 3-flavor ansatz in~\eqnref{expStr3f}.

For the~\figref{instaFlower}\subref{subfig:bubble2}-\subref{subfig:bubble4} diagrams, we need to evaluate diagrams with explicit quark mass insertions. Correspondingly, since the diagram vanishes as $m_q \to 0$ and the explicit quark mass provides a cutoff for the instanton size integration, we use $m_q \Lambda_q^2$ for closing the instanton flower diagram with a quark mass insertion.  The remaining instanton diagrams become
\begin{align}
A_2 &= \frac{v^6}{2} m_u \Lambda_u^2 \cos(\theta_{\pi^0} + \theta_{\eta'} - c_3^G \theta_a ) \ , 
\label{eq:A2instanton}\\
A_3 &= \frac{v^6}{2} m_d \Lambda_d^2 \cos(-\theta_{\pi^0} + \theta_{\eta'} - c_3^G \theta_a ) \ , 
\label{eq:A3instanton}\\
A_4 &= \frac{v^6}{2} m_s \Lambda_s^2 \cos(2\theta_{\eta'} - c_3^G \theta_a ) \ .
\label{eq:A4instanton}
\end{align}
Now the overall axion potential is
\begin{align}
\nonumber
\mathcal{L}
&\supset -m_u v^3 \cos(\theta_{\pi^0} + \theta_{\eta'} + c_2^u \theta_a) - m_d v^3 \cos(-\theta_{\pi^0} + \theta_{\eta'} + c_2^d \theta_a) - \frac{v^9}{4 K^5} \cos(2\theta_{\eta'} - c_3^G \theta_a ) \\
& - \frac{v^6}{2K^5} m_u \Lambda_u^2 \cos(\theta_{\pi^0} +\theta_{\eta'} - c_3^G \theta_a ) - \frac{v^6}{2K^5} m_d \Lambda_d^2 \cos(- \theta_{\pi^0} + \theta_{\eta'} - c_3^G \theta_a ) \ ,
\end{align}
where we neglect the $A_4$ contribution as mentioned previously, since the explicit $m_s$ dependence will give mixing angles of PNGBs suppressed by the strange mass.  This is consistent with neglecting the $\eta$ meson in the PNGB mass matrix.

We can now evaluate the mass squared matrix for the PNGBs by expanding the cosine functions to quadratic order.  We obtain 
\begin{align}
\mathcal{L} &\supset 
\frac{1}{2}\theta_a^2 \left( v^3 ( m_u (c_2^u)^2 + m_d (c_2^d)^2 ) + \left( c_3^G \right)^2 \left( \frac{v^9}{4 K^5} + \frac{v^6}{2 K^5} m_u \Lambda_u^2 + \frac{v^6}{2 K^5} m_d \Lambda_d^2 \right)  \right) \nonumber \\
&+\frac{1}{2} \theta_{\eta'}^2 \left( m_u v^3 + m_d v^3 +  \frac{v^9}{K^5} + \frac{v^6}{2 K^5} m_u \Lambda_u^2 + \frac{v^6}{2 K^5} m_d \Lambda_d^2 \right) \nonumber \\
&+ \frac{1}{2} \theta_{\pi^0}^2 \left( m_u v^3 + m_d v^3 +  \frac{v^6}{2 K^5} m_u \Lambda_u^2 + \frac{v^6}{2 K^5} m_d \Lambda_d^2 \right) \nonumber \\
&+ \theta_a \theta_{\eta'} \left( v^3 ( m_u c_2^u + m_d c_2^d )  - c_3^G \left( \frac{ v^9}{2 K^5} + \frac{v^6}{2 K^5}m_u\Lambda_u^2 + \frac{v^6}{2 K^5} m_d\Lambda_d^2 \right) \right) \nonumber \\
&+\theta_a \theta_{\pi^0}\left( v^3 ( m_u c_2^u - m_d c_2^d ) + c_3^G \left( -\frac{v^6}{2 K^5} m_u \Lambda_u^2 + \frac{v^6}{2 K^5} m_d \Lambda_d^2 \right) \right) \nonumber \\
&+\theta_{\eta'} \theta_{\pi^0} \left( m_u v^3 - m_d v^3 + \frac{v^6}{2 K^5} m_u \Lambda_u^2 - \frac{v^6}{2 K^5} m_d \Lambda_d^2 \right) \ .
\label{eq:thetaexp}
\end{align}

We introduce new compact notation to simplify the discussion,
\begin{align}
m_+ &= m_u + m_d \ , \quad m_- = m_d - m_u \ , \quad m_{c^+} = m_u c_2^u + m_d c_2^d \ , \quad m_{c^-} = m_d c_2^d - m_u c_2^u \ , 
\nonumber \\
\mu &= \dfrac{m_u m_d}{m_u + m_d} \ , \quad \mu L^2 = m_u \Lambda_u^2 + m_d \Lambda_d^2 \ , \quad \Lambda_{\eta'}^4 = \frac{v^9}{4 K^5} \ ,\quad \Lambda_{\text{inst}}^3 = \frac{L^2}{4 K^5}v^6 \ ,
\label{eq:KimNewVariables}
\end{align}
where $\Lambda_{\eta'}$ reflects the instanton contribution to the $\eta'$ meson and $\Lambda_{\text{inst}}$ reflects the instanton contributions from an explicit quark mass.  In this notation, we can write the symmetric PNGB mass squared matrix, $M^2$, as
\begin{align}
\mathcal{L} &=
\frac{1}{2} \left( a \quad \eta' \quad \pi^0 \right)
M^2 \left( a \quad \eta' \quad \pi^0 \right)^T \ , 
\label{eq:massMatrixKim} 
\end{align}
with entries 
\begin{align}
(M^2)_{11} &=
\frac{1}{F_a^2}\left( v^3 ( m_u (c_2^u)^2 + m_d (c_2^d)^2 ) + (c_3^G)^2 \left( \Lambda_{\eta'}^4 + 2\mu \Lambda_{\text{inst}}^3 \right) \right) \ , \nonumber \\
(M^2)_{12} &=
\frac{1}{F_aF_{\eta'}}\left(  m_{c^+} v^3  - c_3^G \left( 2 \Lambda_{\eta'}^4 +  2 \mu \Lambda_{\text{inst}}^3 \right) \right) \ , \nonumber \\
(M^2)_{13} &= 
\frac{- m_{c^-} v^3 }{F_a F_{\pi^0}} \ , \nonumber \\
(M^2)_{22} &= 
\frac{1}{F_{\eta'}^2}(m_+ v^3 + 4\Lambda_{\eta'}^4 + 2\mu  \Lambda_{\text{inst}}^3) \ , \nonumber \\
(M^2)_{23} &=
\frac{-m_- v^3}{F_{\pi^0} F_{\eta'}} \ , \nonumber \\
(M^2)_{33} &= 
\frac{1}{F_{\pi^0}^2}( m_+ v^3 + 2\mu \Lambda_{\text{inst}}^3) \ , 
\label{eq:massMatrixentries}
\end{align}
where we approximate $m_d \Lambda_d^2 - m_u \Lambda_u^2 \approx 0$ from isospin symmetry. 

We diagonalize a generalized form of this matrix in Appendix~\ref{subsec:genMatrix3x3},~\ref{subsec:genMatrix3x3Expansions} taking advantage of an assumed $F_a \gg F_\pi$, $F_{\eta'}$ hierarchy, which affords a robust analytic calculation to account for axion mixing with the $\eta'$ and $\pi^0$ mesons.  For $\pi^0$ and $\eta'$ the mass formulas are
\begin{align}
    m_{\pi^0, \eta'}^2 &= \frac{m_+v^3+4\Lambda_{\eta'}^4+ 2\mu  \Lambda_{\text{inst}}^3}{2F_{\eta'}^2} + \frac{m_+v^3+ 2\mu \Lambda_{\text{inst}}^3}{2F_{\pi^0}^2} \nonumber\\
    &\mp \sqrt{\left(\frac{m_+v^3+4\Lambda_{\eta'}^4+ 2\mu  \Lambda_{\text{inst}}^3}{2F_{\eta'}^2} - \frac{m_+ v^3 + 2\mu \Lambda_{\text{inst}}^3}{2F_{\pi^0}^2} \right)^2 + \frac{m_-^2v^6}{F_{\eta'}^2 F_{\pi^0}^2}} \ , \label{eq:pionEtaMass3x3}
\end{align}
as seen in~\eqnref{pionEtaMass3x3App}. 

The known $m_{\pi^0}$ and $m_{\eta'}$ masses hence constrain $\Lambda_{\eta'}$, $\Lambda_{\text{inst}}$, and $v$ from~\eqnref{pionEtaMass3x3}, but since we have two constraints and three variables, there is not a unique solution.  We adopt the following experimental inputs~\cite{ParticleDataGroup:2020ssz},
\begin{align}
\nonumber
F_{\pi^0} &\approx 131~\text{MeV}, &&  F_{\eta'} \approx 121~\text{MeV}, \\
\nonumber
m_{\pi^0} &= 134.9768(5)~\text{MeV}, && m_{\eta'}= 957.78(6)~\text{MeV}, \\
\nonumber
m_+ &= 6.9^{+1.1}_{-0.3}~\text{MeV}, && \mu=\frac{Zm_+}{(1+Z)^2} =  1.42~\text{MeV}.\\
m_u &= 2.16^{+0.5}_{-0.26}~\text{MeV} \ , && m_d = 4.67^{+0.48}_{-0.17}~\text{MeV} \ ,
\label{eq:values}
\end{align}
which define a solution set of $\Lambda_{\eta'}$, $\Lambda_{\text{inst}}$, and $v$ with a narrow range for $\Lambda_{\eta'}$ given as $239.297 \text{MeV} \lesssim \Lambda_{\eta'} \lesssim 239.306 \text{MeV}$.  We hence fix $\Lambda_{\eta'} = 293.3$~MeV and obtain\footnote{We remark that Ref.~\cite{Kim:2008hd} has $\Lambda_{\eta'} \approx 202 \text{MeV}$ but uses separate convention for $F_{\pi^0}$. According to our convention of $F_{\pi^0}$, the $\Lambda_{\eta'}$ in Ref.~\cite{Kim:2008hd} converts to $\sqrt[4]{2}\cdot 202~\text{MeV} \approx 240$ MeV.  In addition, our numerical results do not change meaningfully when including the quoted experimental uncertainties.}
\begin{align}
v &= 336.3~\text{MeV} \ , \quad \Lambda_{\eta'} = 239.3~\text{MeV} \ , \quad \Lambda_{\text{inst}} =  261.7~\text{MeV} \nonumber \\
\Rightarrow K &= 582.6~\text{MeV} , \quad L = 1289.5~\text{MeV} \ ,
\label{eq:freeValuesSet}
\end{align}
where $K$ is the instanton amplitude from~\eqnref{Ldet2} and $L$ is defined in~\eqnref{KimNewVariables}.  While the numerical results on $\Lambda_{\text{inst}}$ and $v$ depend sensitively on $\Lambda_{\eta'}$ self-consistently with the pion and $\eta'$ mass constraints, we have checked numerically that the resulting axion mass has no significant sensitivity to this variation. 

The full expression for the axion mass eigenvalue from~\eqnref{massMatrixentries} is
\begin{align}
m_a^2F_a^2 &= \left( v^3 ( m_u (c_2^u)^2 + m_d (c_2^d)^2 ) + (c_3^G)^2 \left( \Lambda_{\eta'}^4 + 2\mu \Lambda_{\text{inst}}^3 \right) \right) \nonumber\\
&- \frac{( m_+ v^3 + 2\mu \Lambda_{\text{inst}}^3) \left(  m_{c^+} v^3  - c_3^G \left( 2 \Lambda_{\eta'}^4 +  2 \mu \Lambda_{\text{inst}}^3 \right) \right)^2 }{(m_+ v^3 + 4\Lambda_{\eta'}^4 + 2\mu  \Lambda_{\text{inst}}^3) ( m_+ v^3 + 2\mu \Lambda_{\text{inst}}^3) - (m_- v^3)^2}\nonumber\\
&+ \frac{  2 m_- v^3 \left(  m_{c^+} v^3  - c_3^G \left( 2 \Lambda_{\eta'}^4 +  2 \mu \Lambda_{\text{inst}}^3 \right) \right)  m_{c^-} v^3 }{(m_+ v^3 + 4\Lambda_{\eta'}^4 + 2\mu  \Lambda_{\text{inst}}^3) ( m_+ v^3 + 2\mu \Lambda_{\text{inst}}^3) - (m_- v^3)^2}\nonumber\\
&- \frac{ (m_+ v^3 + 4\Lambda_{\eta'}^4 + 2\mu  \Lambda_{\text{inst}}^3) ( m_{c^-} v^3)^2}{(m_+ v^3 + 4\Lambda_{\eta'}^4 + 2\mu  \Lambda_{\text{inst}}^3) ( m_+ v^3 + 2\mu \Lambda_{\text{inst}}^3) - (m_- v^3)^2} \ , \label{eq:axionMass3x3Extra}
\end{align}
as seen in~\eqnref{axionMass3x3App}. In the case of a KSVZ model~\cite{Kim:1979if, Shifman:1979if}, we can set $c_2^u = c_2^d = 0$ and $c_3^G = 1$, and the axion mass simplifies to
\begin{align}
(m_a^2 F_a^2)^\text{KSVZ} &= \Lambda_{\eta'}^4+2\mu  \Lambda_{\text{inst}}^3-\frac{(2\Lambda_{\eta'}^4+2\mu  \Lambda_{\text{inst}}^3)^2(m_+v^3+ 2\mu  \Lambda_{\text{inst}}^3)}{F_{\pi^0}^2m_{\pi^0}^2F_{\eta'}^2m_{\eta'}^2} \ , \label{eq:KSVZaxionMassEqu} \\
    \Rightarrow m_a^{\text{KSVZ}} &=  8.4 ~\mu\text{eV} \frac{10^{12} ~\text{GeV}}{F_a} \ ,
\label{eq:KSVZaxionMass}
\end{align}
as seen in~\eqnref{KSVZaxionMassApp}.  It is easily checked that without instanton effects, {\it i.e.} in the limit $\Lambda_{\eta'}$, $\Lambda_{\text{inst}} \rightarrow 0$, the axion mass vanishes.

In a DFSZ model~\cite{Dine:1981rt, Zhitnitsky:1980tq}, we have instead $c_3^G = 0$, giving an axion mass
\begin{align}
(m_a^2F_a^2)^\text{DFSZ}&=\frac{4\mu\Lambda_\text{inst}^3(2\Lambda_{\eta'}^4+\mu\Lambda_\text{inst}^3)((c_2^d)^2m_d+(c_2^u)^2m_u)v^3}{F_{\pi^0}^2m_{\pi^0}^2F_{\eta'}^2m_{\eta'}^2}\nonumber\\
&\quad +\frac{4m_um_dv^6 ((c_2^d+c_2^u)^2\Lambda_{\eta'}^4+ ((c_2^d)^2+(c_2^u)^2)\mu\Lambda_\text{inst}^3)}{F_{\pi^0}^2m_{\pi^0}^2F_{\eta'}^2m_{\eta'}^2} \ ,
\label{eq:DFSZaxionMass}
\end{align}
as seen in~\eqnref{DFSZaxionMassApp}. For $c_2^u=c_2^d\equiv c_2$ this simplifies further to
\begin{align}
    (m_a^2F_a^2)^\text{DFSZ}&=2c_2^2(2\Lambda_{\eta'}^4+\mu  \Lambda_{\text{inst}}^3)-\frac{4c_2^2 (2\Lambda_{\eta'}^4+\mu  \Lambda_{\text{inst}}^3)^2(m_+v^3+ 2\mu  \Lambda_{\text{inst}}^3)}{F_{\pi^0}^2m_{\pi^0}^2F_{\eta'}^2m_{\eta'}^2} \ . \nonumber \\
    \Rightarrow m_a^{\text{DFSZ}} &= 15 \mu\text{eV} \frac{10^{12} \text{GeV}}{F_a} \ ,
    \label{eq:DFSZaxionmass2}
\end{align}
where $c_2 = 1$~\cite{Dine:1981rt, Zhitnitsky:1980tq}.  Again, we see that the axion mass vanishes without instanton effects.  These numerical results for the axion masses agree with other published values in the literature~\cite{ParticleDataGroup:2020ssz} at our leading order QCD calculation.  

We note that the mass difference between~\eqnref{KSVZaxionMass} and~\eqnref{DFSZaxionmass2} comes from consistently treating $c_1^i = 0$ for all SM quarks throughout the whole calculation.  If $c_1^i \neq 0$ for some quarks, there is additional kinetic mixing between the axion field and other PNGBs, which would lead to an additional mass mixing effect and is generally ignored in the literature.  We remark that the instanton loops from~\eqnref{A1instanton}-~\eqnref{A4instanton} have been precisely determined in Ref.~\cite{Csaki:2019vte}, but our main focus is the mass mixing of the axion degree of freedom with other PNGBs.

In summary, our generalized calculation to generate and diagonalize the PNGB mass matrix including instanton effects via instanton flower diagrams provides consistent axion masses compared to previous methods.  We will discuss a comparison between our calculation and the historical approach in~\subsecref{comparison}.  We now focus on the PNGB mixing effects on the diphoton coupling. 

\subsection{Axion-Diphoton Coupling}
\label{subsec:axionDiPhotonCoupling}
In this section, we determine the axion-diphoton coupling $G_{a\gamma\gamma}$.  Given the effective Lagrangian~\eqnref{LSM}, the axion-diphoton coupling arises from the gauge contact interactions of $c_3^W$ and $c_3^B$ as well as the loop-induced contributions of the SM fermions.  Consistent with before, we assume $c_1^i = 0$ for the SM quarks and avoid fermion redefinitions.  We remark that~\eqnref{LSM} has an implicit scale assumption where the SM fermions included in the sum modify the corresponding $c_3^G$, $c_3^W$ and $c_3^B$ effective couplings as we renormalize from the electroweak scale to lower scales, which has been extensively addressed in Ref.~\cite{Bauer:2020jbp}, for example.  Correspondingly, we define the contact coupling of photons to the unmixed axion as 
\begin{align}
E \equiv c_3^W + c_3^B + c_3^{\gamma ,\text{ (eff)}},
\label{eq:Etilde}
\end{align}
where we now consider the broken phase of electroweak symmetry and $c_3^{\gamma, \text{ (eff)}}$ encodes the loop effects of all SM quarks with nonzero $c_2^q$ in~\eqref{eq:LSM}, where the loop function and hence $E$ remains $m_a$ dependent.  The gauge couplings of the axion become 
\begin{align}
\mathcal{L} \supset -\frac{a}{F_a}\left(c_3^G \frac{\alpha_s}{8\pi} G\tilde{G} + E \frac{\alpha_e}{8\pi} F\tilde{F}\right) \ .
\label{eq:Lgauge}
\end{align}

The above Lagrangian is not complete, since the mass mixing effects with other PNGBs in low-energy QCD are not yet considered.  We include the mixing contribution from the eigensystem defined in~\eqnref{massMatrixentries}, based on the generalized results in~\eqnref{ev1Sol2}.  Since the PNGB mass eigenstates are defined by
\begin{align}
\mqty(a_m\\\eta'_m\\\pi^0_m) = V^T \mqty(a\\\eta'\\\pi^0),
\end{align}
where $V$ is defined as $\left( \vec{v}_1 |  \vec{v}_2 | \vec{v}_3\right)$ for normalized eigenvectors $\vec{v}_1$, $\vec{v}_2$, and $\vec{v}_3$, we obtain the net electromagnetic coupling for the axion mass eigenstate as a coherent sum from the corresponding "gauge" eigenstate PNGBs, weighted by the eigenvector entries which are in turn approximated in a $\mathcal{O}(1 / F_a)$ expansion:
\begin{align}
\nonumber
\mathcal{L}&\supset  -\frac{1}{4}\left( \left(\frac{\alpha_e}{2\pi F_a} E \right)v_{1,1} +G_{\eta'\gamma\gamma}v_{1,2} + G_{\pi^0\gamma\gamma}v_{1,3}\right) a_m F_{\mu\nu} \tilde{F}^{\mu\nu}\\[12pt]
&\simeq -\frac{1}{4} \left( \frac{\alpha_e}{2\pi F_a} \right) \left( E - \Delta_M \right) a_m F_{\mu\nu} \tilde{F}^{\mu\nu} \ ,\qquad \Delta_M \equiv -\frac{2\pi}{\alpha_e}\left( G_{\eta'\gamma\gamma}(F_a v_{1,2})+ G_{\pi^0\gamma\gamma}(F_a v_{1,3}) \right) \ .
\label{eq:defDeltaM}
\end{align}
Here, $\Delta_M$ defines the correction to the axion-diphoton coupling arising from mixing effects.  We extract $G_{\eta'\gamma\gamma}$ and $G_{\pi^0\gamma\gamma}$ from their respective diphoton decay widths~\cite{ParticleDataGroup:2020ssz}, giving \begin{align}
G_{\eta'\gamma\gamma} &= \sqrt{ \frac{192\pi \Gamma(\eta'\rightarrow\gamma\gamma)}{m_{\eta'}^3} } = 1.72 \cdot 10^{-5} \text{ MeV}^{-1} \ ,\\
G_{\pi^0\gamma\gamma} &= \sqrt{ \frac{64\pi \Gamma(\pi^0\rightarrow\gamma\gamma)}{m_{\pi^0}^3} } = 2.53 \cdot 10^{-5} \text{ MeV}^{-1} .
\end{align}
Using our numerical values from Eq.~\eqref{eq:freeValuesSet}, we obtain
\begin{align}
-\Delta_M &= 1.3 c_3^G + 1.6 c_2^d - 0.8 c_2^u \ ,
\end{align}
which leads to $\Delta_M^{\text{KSVZ}} = -1.3$ and $\Delta_M^{\text{DFSZ}} = -0.8$. 
In contrast to the standard lore, the masses of QCD axions with small size instanton effects can be comparable to or even exceed the electroweak scale, and thus the usual treatment of the electric and color anomaly factors must be amended to "integrate in" the dynamical Standard Model fermions.  Hence, Eq.~\eqref{eq:LSM} serves as a modern axion EFT description since we can self-consistently evaluate the finite fermion mass effects in the loop functions for SM fermions with non-vanishing $c_2^i$.  Nevertheless, we can also reproduce the essential features of the standard QCD axion story using Eq.~\eqref{eq:LSM}.  We first normalize the axion $G \tilde{G}$ coupling by the color anomaly coefficient $c_3^G$, $\tilde{F}_a \equiv F_a / c_3^G$, which leads to the typical rescaling of the diphoton coupling,
\begin{align}
\mathcal{L} &\supset  -\frac{1}{4} \frac{\alpha_e}{2\pi \tilde{F}_a} \left( \frac{E - \Delta_M}{c_3^G} \right) a_m F_{\mu\nu} \tilde{F}^{\mu\nu} \equiv
-\frac{1}{4} G_{a\gamma\gamma} a_m F_{\mu\nu} \tilde{F}^{\mu\nu}\ .
\label{eq:elMagLagrEigensystem}
\end{align}

\begin{figure}[htb]
\centering
\includegraphics[width=\textwidth]{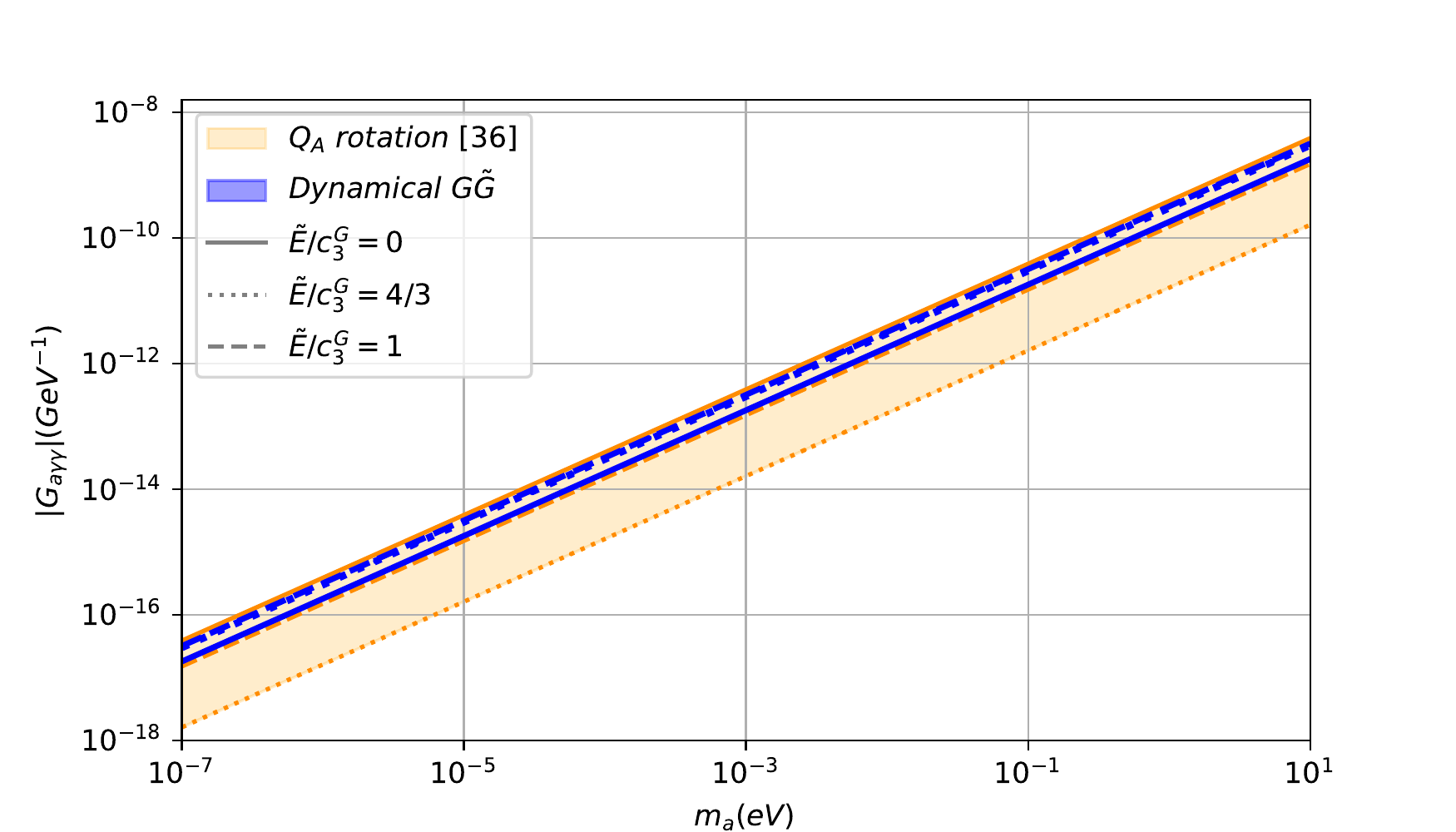}
\caption[ Comparison of the axion-bands]{ Crosscheck of the typical QCD axion band~\cite{ParticleDataGroup:2020ssz, diCortona:2015ldu} (blue) and our instanton-flower diagram results (orange) in the $(m_a,G_{a\gamma\gamma})$ plane.  We relate our $\tilde{E} / c_3^G$ parameter to the canonical $E / N$ parameter from the canonical DFSZ and KSVZ models via $\tilde{E} / c_3^G = 0.5 (E / N)$, and we show three explicit choices of $\tilde{E} / c_3^G = 0$, $1$, and $4/3$. 
\label{fig:cChiCompAxBand}}
\end{figure}

We show our results for the typical QCD axion band compared to the standard QCD axion band from Ref.~\cite{ParticleDataGroup:2020ssz} based on Ref.~\cite{diCortona:2015ldu}.  While our results do not show significant discrepancies in the $(m_a, G_{a\gamma \gamma})$-parameter space, as shown in Fig.~\ref{fig:cChiCompAxBand}, we attribute the residual differences to our lack of next-to-leading order corrections and our missing contribution of the $\eta$-meson, which should correct our derived $\tilde{G}_{a\gamma \gamma}$ couplings by $\mathcal{O}(30\%)$ when $c_3^G \simeq 1$.\footnote{We remark that our notation uses the fraction $\tilde{E} / c_3^G = 0.5 (E / N)$ for the effective fermion contribution to the diphoton coupling via anomalies, where $E / N$ is the notation from Ref.~\cite{ParticleDataGroup:2020ssz}.}


\subsection{Cross-check to earlier results on axion mass derivation}
\label{subsec:comparison}

We can cross-check our results on the axion mass, where we treat the axion and other PNGBs as phases of the $G \tilde{G}$ operator, to the traditional method established in Ref.~\cite{Georgi:1986df} by Georgi, Kaplan and Randall.  We again emphasize that our framework is originated from Ref.~\cite{Kim:2006aq, Kim:2008hd}: the main benefit of our framework is that we can readily calculate effects from SSIs, where the traditional approach fails.


Starting with the general Lagrangian in Eq.~\ref{eq:LSM}, the traditional approach by Ref.~\cite{Georgi:1986df} rotates the $c_3^G G \tilde{G}$ term into the quark couplings via the global phase rotation
\begin{align}
q\rightarrow \exp(-i\frac{c_3^Ga}{F_a} (Q_V + Q_A \gamma_5) ) q \ , \qquad \Tr(Q_A)=1/2 \ , \qquad q = \mqty(u\\d\\s) \ ,
\label{eq:Georgirot}
\end{align}
where $Q_V$ can be taken diagonal but $Q_A$ is arbitrary except of its trace. 
This results in the following Lagrangian,
\begin{align}
\label{eq:lowEnergyLagr1}
\mathcal{L} &= \frac{1}{2}\partial_\mu a \partial^\mu a + \left( \sum\limits_{i = 1}^{N_f} -\bar{q}_R^i \tilde{M} q_L^i +\frac{\partial_\mu a}{F_a} \bar{q}^i \gamma_\mu \left( Q_V + (c_1^i + Q_A) \gamma_5 \right) q^i + \text{ h.c.} \right) -\frac{a}{F_a} \frac{\alpha_e}{8\pi} c_3^\gamma F\tilde{F} \ ,
\end{align}
where 
\begin{align}
\tilde{M} &= \exp( 
\dfrac{i a (c_2^q + c_3^G) Q_A }{F_a} ) M \exp( \dfrac{i a (c_2^q + c_3^G) Q_A }{F_a}) \ ,
\end{align}
for $M$ the diagonalized quark mass matrix, and 
\begin{align}
c_3^\gamma = c_3^W + c_3^B - 4\Tr(Q_AQ_EQ_E) \equiv c_3^W + c_3^B - c_\chi, \qquad Q_E=\mqty(\dmat{2/3,-1/3,2/3}) \ ,
\end{align}
where $Q_E$ is the electric charge matrix of the quarks and  $c_{\chi}$ emerges from effects of chiral symmetry breaking.  We next match the operators in Eq.~\ref{eq:lowEnergyLagr1} to axion-dependent terms in the effective chiral Lagrangian (ChEFT).  We use the mass matrix $\tilde{M}$ to encode the axion interactions in the ChEFT-Lagrangian,
\begin{align}
\label{eq:LCheft}
\mathcal{L}_\text{ChEFT} = \frac{1}{4}F_\pi^2 \Tr(D^\mu \sigma D_\mu \sigma^\dag)+\frac{1}{2} F_\pi^2 \mu \Tr(\sigma \tilde{M}) + \text{h.c.} \ ,
\end{align}
where $\sigma = \exp(2i \pi^a t^a/F_\pi)$ transforms under $SU(3)_L \times SU(3)_R$, $\mu$ is related to pseudoscalar masses by the Gell-Mann-Oakes-Renner relation, and $F_\pi\simeq 131$~MeV. Next, we diagonalize the mass term in Eq.~\ref{eq:LCheft}.  This is done by a specific choice of $Q_A$, 
\begin{align}
\label{eq:qa}
Q_A = \frac{1}{2} M^{-1}/\Tr(M^{-1}) \ .
\end{align}
Expanding the exponential factors in $\tilde{M}$ leads to
\begin{align}
\mathcal{L}_{\text{ChEFT}} &\supset \frac{1}{2} \left( \frac{a}{F_a} \right)^2 \Tr(\{\{M, Q_A\}, Q_A\} \sigma) + \text{ h.c.} \ ,
\end{align}
where non-diagonal terms successfully vanish for the given choice of $Q_A$. Therefore, no further diagonalization of the mass matrix is needed and the axion mass turns out as 
\begin{align}
m_a^2 = \frac{1}{F_a^2} \frac{m_\pi^2 F_\pi^2}{(m_u + m_d) \Tr(M^{-1})} \ .
\label{eq:simpleAxionMass}
\end{align}

We emphasize that this solution for the axion mass is only valid for the case where the axion potential is entirely determined by QCD effects. Note that the axion potential from QCD instantons has not been explicitly calculated but instead only included through the mass mixing with the other pseudoscalars of the theory, in particular the pion.  In particular, the calculation of QCD instanton effects has been circumvented by choosing the axion basis rotation in Eq.~\ref{eq:Georgirot}.  

We emphasize that the axion basis achieved by Eq.~\ref{eq:Georgirot} does not guarantee the absence of small-size instanton effects, since a non-trivial embedding of $SU(3)_c$ can generate a larger PNGB mass matrix necessitating more axial rotations of colored fields, which cannot be assumed for models with bifundamental matter.
Hence for the general case, it is necessary to integrate out all of the non-Abelian gauge fields and to determine all of the mass mixings with other PNGBs explicitly.

\subsubsection{Comparison of the two methods}

Now that we have concluded with the traditional derivation of axion mass based on Ref.~\cite{Georgi:1986df}, we can compare to our framework based on Ref.~\cite{Kim:2006aq, Kim:2008hd}.  These two methods are related by an axial transformation of the quark fields, Eq.~\ref{eq:Georgirot}, which was used in Ref.~\cite{Georgi:1986df} to remove the $\theta G \tilde{G}$ operator.  Clearly, such a basis transformation cannot have an effect on physical observables.  In our framework, however, the $\eta'$ meson correctly receives the QCD instanton contribution to its mass while this is neglected in the traditional approach.  We will now demonstrate that our mass eigenstates reproduce the results of Ref.~\cite{Georgi:1986df} in this limit.

We can most easily compare our results on $\pi^0$, $\eta'$, and axion masses in~\eqnref{pionEtaMass3x3} and~\eqnref{axionMass3x3Extra} to established results by expanding in isospin symmetry breaking as well as the instanton effect on the $\eta'$ mass.  For simplicity, we will focus on the KSVZ case in~\eqnref{KSVZaxionMassEqu} with $c_2^u = c_2^d = 0$ and normalize $c_3^G = 1$, and we can define
\begin{align}
\Delta_m^2 \equiv \frac{1}{F_{\eta'}^2} (m_+v^3 + 4\Lambda_{\eta'}^4 + 2\mu  \Lambda_{\text{inst}}^3) - \frac{1}{F_{\pi^0}^2}( m_+v^3 + 2\mu  \Lambda_{\text{inst}}^3) \ .
\end{align}
For small instanton amplitudes, $K \ll v$, or, alternately, equal decay constants $F_{\eta'}/F_{\pi^0} \to 1$, we get $\Delta_m^2 \to 4 \Lambda_{\eta'}^4 / F_{\eta'}^2$. 

We can now equivalently take the isospin limit or the no-instanton effect on $\eta'$ limit by expanding in the ratio $4 m_-^2 v^6 / (\Delta_m^4 F_{\pi^0}^2 F_{\eta'}^2)$.  As shown in the appendix in Eqs.~\ref{eq:etaMassexpansionApp} and~\ref{eq:KimPionMassApp}, our mass eigenvalues become
\begin{align}
m_{\eta'}^2 &= \frac{m_+v^3 + 4\Lambda_{\eta'}^4 + 2\mu  \Lambda_{\text{inst}}^3}{F_{\eta'}^2} + \frac{\Delta_m^2}{2} \sum_{k=1}^{\infty} {1/2\choose k} \left( \frac{4m_-^2v^6}{ \Delta_m^4 F_{\pi^0}^2 F_{\eta'}^2} \right)^k 
\label{eq:etaMassexpansion} \ , \\
m_{\pi^0}^2 &= \frac{m_+v^3 + 2\mu \Lambda_{\text{inst}}^3}{F_{\pi^0}^2} -\frac{\Delta_m^2}{2} \sum_{k=1}^{\infty} {1/2\choose k} \left( \frac{4 m_-^2 v^6}{\Delta_m^4 F_{\pi^0}^2 F_{\eta'}^2} \right)^k  \label{eq:KimPionMass} \ .
\end{align}
Thus, the corrections in $m_{\eta'}^2$ and $m_{\pi^0}^2$ vanish in the simultaneous limit $m_-v^3/\Lambda_{\eta'}^4 \to 0$, and the first terms reproduce the known results for the $\eta'$ and $\pi^0$ masses from the solution of the $U(1)$ problem~\cite{tHooft:1986ooh} and the Gell-Mann-Oakes-Renner relation~\cite{Gell-Mann:1968hlm}.

We now expand the axion mass in the limit $\Lambda_{\eta'} \to \infty$, as shown in \eqnref{axionExpansionApp}:
\begin{align}
m_a^2 F_a^2 &= \frac{(m_+v^3 + 2\mu \Lambda_{\text{inst}}^3)^2 - m_-^2 v^6}{4(m_+v^3 + 2\mu \Lambda_{\text{inst}}^3)} \nonumber\\
&-\frac{(4\mu^2 \Lambda_{\text{inst}}^6 + m_+^2 v^6 - m_-^2 v^6)^2}{16 \Lambda_{\eta'}^4 (m_+v^3 + 2\mu \Lambda_{\text{inst}}^3)^2} \sum_{k=0}^{\infty} \left( \frac{m_-^2 v^6 - (m_+v^3 + 2\mu \Lambda_{\text{inst}}^3)^2}{ 4\Lambda_{\eta'}^4 (m_+v^3 + 2\mu \Lambda_{\text{inst}}^3)} \right)^k \ .
\end{align}
Dropping all orders of $1 / ( \Lambda_{\eta'}^4)$ and using the leading order approximation for the pion mass leads to the result in Ref.~\cite{Kim:2008hd},
\begin{align}
m_a^2 F_a^2 &= \frac{m_{\pi^0}^4 F_{\pi^0}^4 - m_-^2 v^6}{4 m_ {\pi^0}^2 F_{\pi^0}^2} = \frac{Z}{(1+Z)^2} m_{\pi^0}^2 F_{\pi^0}^2 \left( 1 + \frac{m_-^2}{m_+} \frac{\Lambda_{\text{inst}}^3 (m_ + v^3 + \mu  \Lambda_{\text{inst}}^3)}{ m_{\pi^0}^4 F_{\pi^0}^4} \right)\ ,
\end{align}
which coincides with Eq.~\ref{eq:simpleAxionMass} and includes a correction from instanton effects from our more general calculation.  Thus, we have demonstrated that our method is consistent with previous derivations of axion properties in the case where no SSI effects take place. 
We will now use our method on more recent models that involve SSI effects, where the traditional approach fails.

\section{Axion Mass and Coupling Enhancemensts from Small-size Instanton Effects}
\label{sec:enhancement}

In this section we discuss the color unification model published in Ref.~\cite{Gaillard:2018xgk} as an exemplary model demonstrating the impact of SSI effects on the axion mass and diphoton coupling.  The color unification model introduces a new, high-quality, composite axion as well as a hidden $\eta_d$ exotic meson, both of which experience mass mixing with SM mesons and SSI effects.  The corresponding axion masses and diphoton couplings are readily calculated in our framework.  In particular, we improve on the results of Ref.~\cite{Gaillard:2018xgk} by including further mixing effects between the exotic PNGBs and separate SM mesons.  We will summarize the current constraints on these SSI-enhanced axions in the $(m_a, G_{a \gamma \gamma})$-plane.


\subsection{Dynamical axion from color unification}
\label{subsec:dynAxCol}



The main idea of the color unification mechanism in Ref.~\cite{Gaillard:2018xgk} is to embed $SU(3)_c$ into another confining $SU(6)$ gauge group with a massless fermion $Q$ at high energies that solves all Strong CP-problems at lower scales. A possible non-zero $\theta_6$ parameter of the $SU(6)$ gauge group is unphysical in the UV due to the massless colored $Q$ fermion, and the 't Hooft anomaly matching condition subsequently guarantees that in the IR $\bar{\theta}$ is unphysical.
The relevant PQ-symmetry for $\theta_6$ must arise from the flavor symmetry of exotic quarks in the UV and is broken by their quark condensates, resulting in a high quality composite axion in the IR. It is important that the anomalous symmetries in the UV significantly affect the masses of the composite states left in the IR, in analogy to $\eta'$ meson for the SM $U(1)$ problem.  One complication of the model is that the SM quarks must be embedded into $SU(6)$ multiplets leading to new exotic electroweak doublets that break the SM electroweak symmetry heavily.  The authors of Ref.~\cite{Gaillard:2018xgk} add an additional $SU(3')$ gauge group and a new bifundamental scalar $\Delta$ such that $\Delta$ spontaneously breaks the product groups at a scale $\Lambda_{\text{CUT}}$ and gives mass to the exotic fermions charged under $SU(3')$, decoupling them from the SM.  Then, the diagonal subgroup $SU(3)_{\text{diag}}$ between $SU(3')$ and $SU(6)$ confines at the scale $v_{\text{diag}}$, leading to the symmetry breaking pattern
\begin{align}
SU(6)\times SU(3') \xrightarrow{\Lambda_{\text{CUT}} } SU(3)_c \times SU(3)_{\text{diag}} \xrightarrow{ v_{\text{diag}} } SU(3)_c \ , \label{eq:symmBreakingPattern}
\end{align}
with $\Lambda_{\text{CUT}}\gg v_{\text{diag}}\gg 246$~GeV.  

An additional complication arises from a possible $\theta'$ term in the $SU(3')$ gauge theory that could spoil the resolution of the $\bar{\theta}$ parameter. To resolve this issue, the authors of Ref.~\cite{Gaillard:2018xgk} consider two possible model extensions, M1 and M2, that relax $\theta'$ to 0, as summarized in~\tableref{GaillardFieldContent}.  The M1 option is a KSVZ-like solution that adds an additional massless fermion $q$ charged under $SU(3')$.  The M2 option is a DFSZ-like solution that adds an additional bifundamental scalar $\Delta_2$ charged under $SU(3')$ and $SU(6)$.

\begin{table}[tb]
\centering
\begin{subfigure}{0.48\textwidth}
\centering
\begin{tabular}{c|c|c||c|c}
&$SU(3)_{\text{diag}}$&$SU(3)_c$&$SU(3')$&$SU(6)$\\
\hline
$Q$ & $\bar{\square} $ & $\square $ & 1 & 20\\
$q$ & $\square $ & 1 & $\square$ & 1
\vspace{12pt}
\end{tabular}
    \caption{}
    \label{subtab:fieldContentM1}
\end{subfigure}
\hfill
\begin{subfigure}{0.48\textwidth}
\centering
\begin{tabular}{c|c|c||c|c}
&$SU(3)_{\text{diag}}$&$SU(3)_c$&$SU(3')$&$SU(6)$\\
\hline
$Q$ & $\bar{\square} $ & $\square $ & 1  & 20\\
$\Delta_2$ & - & - & $\bar{\square}$ & $\square$
\vspace{12pt}
\end{tabular}
    \caption{}
    \label{subtab:fieldContentM2}
\end{subfigure}
    \caption{New field content in the two models (a) M1, KSVZ-like  and (b) M2, DFSZ-like used to solve the $\theta'$ parameter in the $SU(6) \times SU(3)'$ color unification model from Ref.~\cite{Gaillard:2018xgk}.  Both $Q$ and $q$ are massless and are part of the exotic quark condensate, while $\Delta_2$ has a vev $\langle \Delta \rangle  = \Lambda_{\text{CUT}}$.  Compared to the original nomenclature in Ref.~\cite{Gaillard:2018xgk}, we identify $Q=\psi_L$ and $q=\chi$.}
\label{table:GaillardFieldContent}
\end{table}

Overall, the color unification model exhibits two gauge groups $SU(3)_{\text{diag}}$ and $SU(3)_c$ that confine at two separate scales, and the structure of the embedding of $SU(3)_c$ into $SU(6)$ ensures the strong CP problem is solved resulting in a composite axion.  All exotic colored states can be decoupled by raising the confinement scale of $SU(3)_{\text{diag}}$.

\subsubsection{Axion and axieta masses for the M1 variant of the color unification model}

Below the scale $\Lambda_{\text{CUT}}$ and above $SU(3)_{\text{diag}}$ confinement, the relevant Lagrangian for the M1 variant is
\begin{align}
    \mathcal{L} &\supset \bar{Q}_{I,i} \left( i \delta_{IJ} \ \delta_{ij} \ \slashed{\partial} - g_{\text{diag}}\   T_{IJ}^A \slashed{A}_{\text{diag}}^A \delta_{ij} 
    - g_s\  \delta_{IJ} T_{ij}^a \slashed{A}^a \right) Q_{J,j} \nonumber \\
    &  + \bar{q}_{I,i'} \left( i \delta_{IJ}\ \delta_{i'j'} \slashed{\partial} - g_{\text{diag}} \ T_{IJ}^A \slashed{A}_{\text{diag}}^A \delta_{i'j'} - g'\ \delta_{IJ} T_{i'j'}^b \slashed{A}^{\prime, b} \right) q_{J,j'} \nonumber\\
    & + \theta_{\text{diag}} \frac{\alpha_{\text{diag}}}{8\pi} G_{\text{diag}} \tilde{G}_{\text{diag}} + \bar{\theta} \frac{\alpha_s}{8\pi} G \tilde{G} + \theta' \frac{\alpha'}{8\pi} G' \tilde{G}' + \frac{\left( g' \right)^{2} \Lambda_\text{CUT}^2}{2} A'_\mu A^{\prime \mu} \ , 
    \label{eq:LagrCompositeAxionAboveConfinement}
\end{align}
where $(A_{\text{diag}})_\mu$, $A_\mu$ and $A'_\mu$ and $G_{\text{diag}}$, $G$, and $G'$ are the gluon fields and field strengths of $SU(3)_{\text{diag}}$, $SU(3)_c$ and $SU(3')$, $T$ are the Gell-Mann matrices with octet representation indices $A$, $a$, and $b$, and the remaining $i$, $j$, $I$, $J$, $i'$, and $j'$ are color triplet indices.  At a high scale, $g_s \sim 0$ such that there is a $U(4)\times U(4)$ flavor symmetry of the fields $Q_{I1}$, $Q_{I2}$, $Q_{I3}$ and $q_I$. The Lagrangian in \eqnref{LagrCompositeAxionAboveConfinement} mimics the classical composite axion Lagrangian in Ref.~\cite{Kim:1984pt} except for the fact that $q$ has an additional color charge under $SU(3')$ and additional instanton effects arise.

The exotic quark condensate $\langle \bar{Q}_{1}Q_{1}\rangle = \cdots = \langle \bar{q}q \rangle \equiv v_{\text{diag}}^3$ breaks the flavor symmetry $ U(4)\times U(4) = SU(4)_L\times SU(4)_R\times U(1)_V\times U(1)_A$ down to $SU(4)_{\text{iso}}\times U(1)_V\times U(1)_A$, where $U(1)_A$ is broken by the axicolor anomaly. According to the Goldstone theorem we have 16 pNGBs in the broken phase. Fourteen of them are colored exotic mesons with masses $\sim v_{\text{diag}}^2$. We focus on the two color singlet particles in the spectrum that are axion-like particles. We label one as the composite axion $a$ and the other as $\eta_d$, in analogy to the $\eta'$ in QCD.

To calculate the instanton effects on the PNGB masses, we adopt the framework from~\subsecref{method} for the Lagrangian in~\eqnref{LagrCompositeAxionAboveConfinement}.  Namely, we replace the $G_{\text{diag}} \tilde{G}_{\text{diag}}$, $G \tilde{G}$, and $G' \tilde{G}'$ operators by their corresponding 't Hooft determinantal operators. The low-energy QCD mesons are included according to~\eqnref{expUp3f},~\eqnref{expDown3f}, and~\eqnref{expStr3f}, whereas the composite axion field $a$ and exotic axieta $\eta_d$ are included as angular field excitations around the exotic quark condensate $v_{\text{diag}}^3$ as 
\begin{align}
\bar{Q}_L Q_R &\approx \abs{\langle \bar{Q}_L Q_R \rangle}\exp(i \frac{\sqrt{6}a}{F_a}) = \frac{v_{\text{diag}}^3}{2} \exp(i \frac{\sqrt{6}a}{F_a})\ , \text{ and} \label{eq:BigQcondensate}\\
\bar{q}_Lq_R&\approx \abs{\langle \bar{q}_L q_R\rangle} \exp(i \frac{2 \eta_d}{F_a}) = \frac{v_{\text{diag}}^3}{2} \exp(i \frac{2 \eta_d}{F_a}) \ , \label{eq:SmallQcondensate}
\end{align}
where the factors $\sqrt{6}$ and $2$ are from the corresponding flavor symmetry generator (which coincide with the color anomaly prefactors~\cite{Gaillard:2018xgk}).

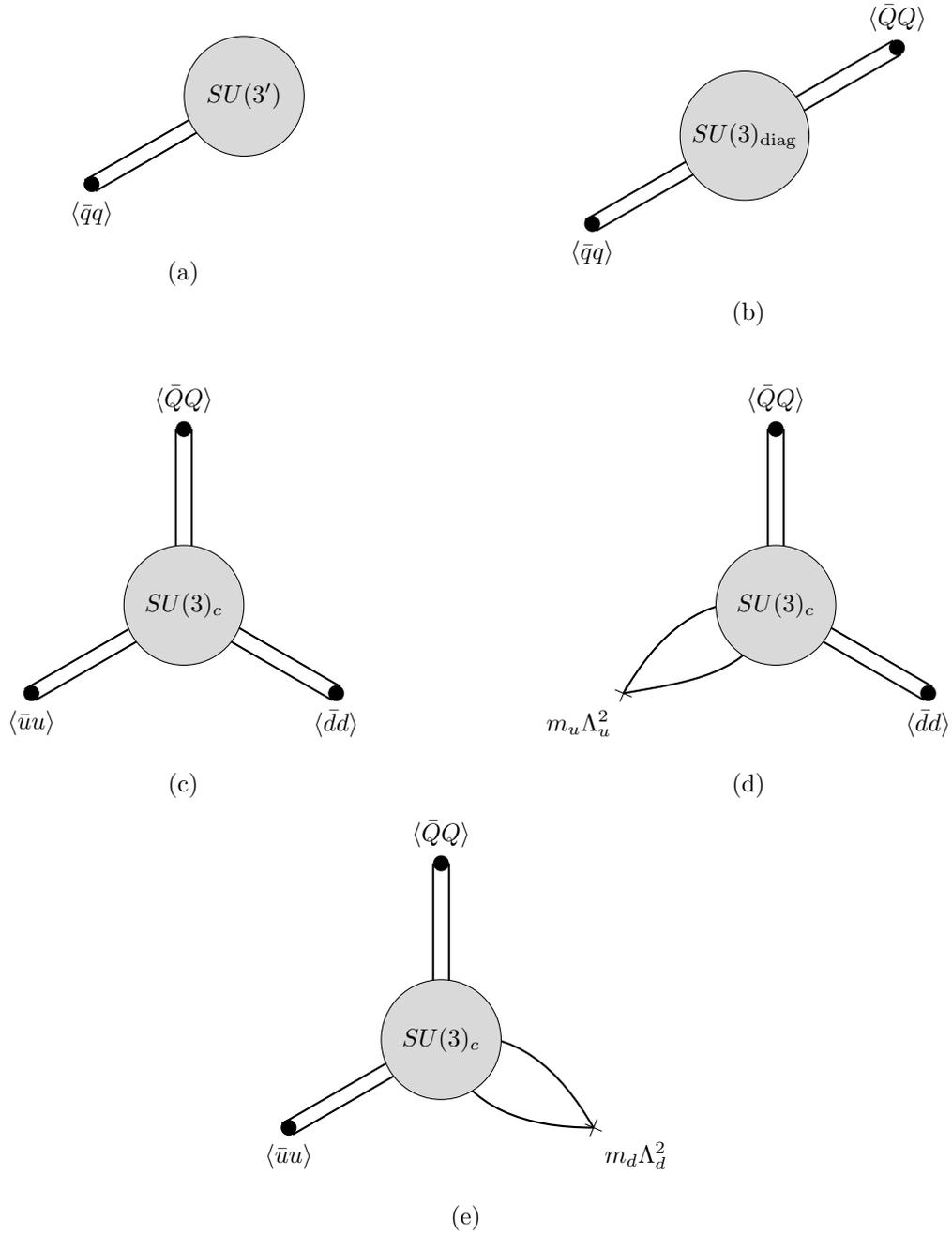
\begin{figure}[tb!]
\centering
\begin{subfigure}{0.48\textwidth}
\centering
\begin{tikzpicture}
\node[circle,draw, minimum size = 1.7cm,color = black, fill = gray!30] (center) at  (0,0) {$SU(3')$};
\begin{scope}[on background layer]
\node[circle,draw,inner sep =0pt, minimum size = 6pt,color = white] (helper) at  (0,0){};
\node[circle,draw,inner sep =0pt, minimum size = 6pt,color = black,fill=black, label = below:$\langle \bar{q}q\rangle$] (A) at (-2.16,-1.25){};
\draw[thick] (helper.-60) -- (A.-60);
\draw[thick] (helper.150) -- (A.150);
\end{scope}
\end{tikzpicture}
\caption{\label{subfig:fig4_bubble1}}
\end{subfigure}
\vspace{15pt}
\begin{subfigure}{0.48\textwidth}
\centering
\begin{tikzpicture}
\node[circle,draw, minimum size = 1.7cm,color = black, fill = gray!30] (center) at  (0,0) {$SU(3)_{\text{diag}}$};
\begin{scope}[on background layer]
\node[circle,draw,inner sep =0pt, minimum size = 6pt,color = white] (helper) at  (0,0){};
\node[circle,draw,inner sep =0pt, minimum size = 6pt,color = black,fill=black, label = below:$\langle \bar{q}q\rangle$] (A) at (-2.16,-1.25){};
\node[circle,draw,inner sep =0pt, minimum size = 6pt,color = black,fill=black, label = above:$\langle \bar{Q}Q\rangle$] (B) at (2.16,1.25){};
\draw[thick] (helper.-60) -- (A.-60);
\draw[thick] (helper.150) -- (A.150);
\draw[thick] (helper.-60) -- (B.-60);
\draw[thick] (helper.120) -- (B.120);
\end{scope}
\end{tikzpicture}
\caption{\label{subfig:fig4_bubble2}}
\end{subfigure}
\vspace{0pt}
\begin{subfigure}{0.48\textwidth}
\centering
\begin{tikzpicture}
\node[circle,draw, minimum size = 1.7cm,color = black, fill = gray!30] (center) at  (0,0) {$SU(3)_c$};
\begin{scope}[on background layer]
\node[circle,draw,inner sep =0pt, minimum size = 6pt,color = white] (helper) at  (0,0){};
\node[circle,draw,inner sep =0pt, minimum size = 6pt,color = black,fill=black, label = below:$\langle \bar{u}u\rangle$] (A) at (-2.16,-1.25){};
\node[circle,draw, ,inner sep =0pt, minimum size = 6pt,color = black,fill=black, label = below:$\langle \bar{d}d\rangle$] (B) at (2.16,-1.25){};
\node[circle,draw,inner sep =0pt, minimum size = 6pt,color = black,fill=black, label = above:$\langle \bar{Q}Q\rangle$] (C) at (0,2.5){};
\draw[thick] (helper.-60) -- (A.-60);
\draw[thick] (helper.150) -- (A.150);
\draw[thick] (helper.60) -- (B.60);
\draw[thick] (helper.240) -- (B.240);
\draw[thick] (helper.0) -- (C.0);
\draw[thick] (helper.180) -- (C.180);
\end{scope}
\end{tikzpicture}
\caption{\label{subfig:fig4_bubble3}}
\end{subfigure}
\begin{subfigure}{0.48\textwidth}
\centering
\begin{tikzpicture}
\node[circle,draw, minimum size = 1.7cm,color = black, fill = gray!30] (center) at  (0,0) {$SU(3)_c$};
\begin{scope}[on background layer]
\node[circle,draw,inner sep =0pt, minimum size = 6pt,color = white] (helper) at  (0,0){};
\coordinate (A) at (-2.16,-1.25){};
\node[cross=3,rotate=22,label = below left:$m_u\Lambda_u^2$] at (A){};
\node[circle,draw, ,inner sep =0pt, minimum size = 6pt,color = black,fill=black, label = below:$\langle \bar{d}d\rangle$] (B) at (2.16,-1.25){};
\node[circle,draw,inner sep =0pt, minimum size = 6pt,color = black,fill=black, label = above:$\langle \bar{Q}Q\rangle$] (C) at (0,2.5){};
\draw [thick] (helper) to [out=250,in=10] (A);
\draw [thick] (helper) to [out=170,in=60](A);
\draw[thick] (helper.60) -- (B.60);
\draw[thick] (helper.240) -- (B.240);
\draw[thick] (helper.0) -- (C.0);
\draw[thick] (helper.180) -- (C.180);
\end{scope}
\end{tikzpicture}
\caption{\label{subfig:fig4_bubble4}}
\end{subfigure}
\begin{subfigure}{0.48\textwidth}
\centering
\begin{tikzpicture}
\node[circle,draw, minimum size = 1.7cm,color = black, fill = gray!30] (center) at  (0,0) {$SU(3)_c$};
\begin{scope}[on background layer]
\node[circle,draw,inner sep =0pt, minimum size = 6pt,color = white] (helper) at  (0,0){};
\node[circle,draw,inner sep =0pt, minimum size = 6pt,color = black,fill=black, label = below:$\langle \bar{u}u\rangle$] (A) at (-2.16,-1.25){};
\coordinate (B) at (2.16,-1.25){};
\node[cross=3,rotate=22,label = below:$m_d\Lambda_d^2$] at (B){};
\node[circle,draw,inner sep =0pt, minimum size = 6pt,color = black,fill=black, label = above:$\langle \bar{Q}Q\rangle$] (C) at (0,2.5){};
\draw[thick] (helper.-60) -- (A.-60);
\draw[thick] (helper.150) -- (A.150);
\draw [thick] (helper) to [out=370,in=120] (B);
\draw [thick] (helper) to [out=290,in=180](B);
\draw[thick] (helper.0) -- (C.0);
\draw[thick] (helper.180) -- (C.180);
\end{scope}
\end{tikzpicture}
\caption{\label{subfig:fig4_bubble5}}
\end{subfigure}
\caption[Instanton diagrams for the first considered model]{Instanton diagrams for the M1 model variant of the color unification model from Ref.~\cite{Gaillard:2018xgk}, where $Q$ and $q$ are exotic quarks from Table~\ref{table:GaillardFieldContent}.   We distinguish between instantons from different non-Abelian gauge groups, and quark bilinears are organized according to chirality flipping insertions. Diagrams~(\subref{subfig:fig4_bubble1})$-$(\subref{subfig:fig4_bubble3}) use quark condensates, denoted by $\bullet$, while diagrams~(\subref{subfig:fig4_bubble4}) and (\subref{subfig:fig4_bubble5}) use one insertion of a light quark mass, symbolized by~$\times$.
\label{fig:instaFlowersGai}}
\end{figure}

The resulting axion potential relaxes all $\theta$ parameters, including $\bar{\theta}$ of the SM, thereby solving the Strong CP problem~\cite{Gaillard:2018xgk}. Hence, in the following we look at the excitations around $\langle \eta_d \rangle$ and $\langle a \rangle$.  To determine the axion potential we draw the relevant instanton diagrams, which are shown in~\figref{instaFlowersGai}. The instanton contributions result from three different gauge symmetries and have the form
\begin{align}
\mathcal{L}_{\text{det}}=-K' A_1 - \frac{1}{K_{\text{diag}}^2} A_2 - \frac{1}{K^8} (A_3 + A_4 + A_5) \ ,
\end{align}
where $A_1$ and $A_2$ come from the included 't Hooft determinantal operators for $G' \tilde{G}'$ and $G_{\text{diag}} \tilde{G}_{\text{diag}}$, respectively, while $A_3$, $A_4$ and $A_5$ correspond to $SU(3)_c$ instantons. We remark that the instanton amplitude $K'$ can be large since the coupling of the $SU(3')$ gauge group can be set arbitrarily large. 

The instanton diagram in Fig.~\ref{subfig:fig4_bubble1} is the non-trivial contribution to the axion mass endowed with small-size instanton effects.  Using~\eqnref{expUp3f}$-$~\eqnref{expStr3f} as well as~\eqnref{BigQcondensate}$-$\eqnref{SmallQcondensate}, the amplitudes are
\begin{align}
A_1 &= v_{\text{diag}}^3 \cos(\frac{2 \eta_d}{F_a}) \ , \label{eq:instantonPotentialGaillardA1}\\
A_2 &= \frac{v_{\text{diag}}^6}{2} \cos(\frac{2\eta_d}{F_a} + \frac{\sqrt{6} a}{F_a}) \ , \label{eq:instantonPotentialGaillardA2}\\
A_3 &= \frac{v_{\text{diag}}^3 v^9}{4} \cos( \frac{\sqrt{6} a}{F_a} + 2\frac{\eta'}{F_{\eta'}}) \ ,\\
A_4 &= \frac{v_{\text{diag}}^3 v^6}{2} m_u \Lambda_u^2 \cos( \frac{\sqrt{6} a}{F_a} + \frac{\eta'}{F_{\eta'}} + \frac{\pi^0}{F_{\pi^0}}) \ , \\
A_5 &= \frac{v_{\text{diag}}^3 v^6}{2} m_d \Lambda_d^2 \cos( \frac{\sqrt{6} a}{F_a} + \frac{\eta'}{F_{\eta'}} - \frac{\pi^0}{F_{\pi^0}})\ .
\end{align}
The resulting axion potential is 
\begin{align}
\mathcal{L}\supset & - K' v_{\text{diag}}^3 \cos(\frac{2 \eta_d}{F_a}) 
- \frac{v_{\text{diag}}^6}{2 K_{\text{diag}}^2} \cos(\frac{2 \eta_d}{F_a}+\frac{\sqrt{6} a}{F_a})
\nonumber \\
& - m_u v^3 \cos(\frac{\pi^0}{F_{\pi^0}}+\frac{\eta'}{F_{\eta'}}) 
- m_dv^3\cos(-\frac{\pi^0}{F_{\pi^0}}+\frac{\eta'}{F_{\eta'}}) - \frac{v_{\text{diag}}^3 v^9}{4 K^8} \cos( \frac{\sqrt{6} a}{F_a} + 2\frac{\eta'}{F_{\eta'}})
\nonumber \\
& - \frac{v_{\text{diag}}^3v^6m_u\Lambda_u^2}{2 K^8} \cos(\frac{\sqrt{6} a}{F_a} + \frac{\eta'}{F_{\eta'}}+\frac{\pi^0}{F_{\pi^0}}) 
- \frac{v_{\text{diag}}^3v^6m_d\Lambda_d^2}{2 K^8} \cos(\frac{\sqrt{6} a}{F_a} + \frac{\eta'}{F_{\eta'}} - \frac{\pi^0}{F_{\pi^0}}) \ . \label{eq:axionPotentialGaillard}
\end{align}
We evaluate the mass matrix by expanding the cosine to second order. We obtain
\begin{align}
\mathcal{L} &\supset \frac{1}{2} \left( \frac{\eta_d}{F_a} \right)^2 
4 \left( K' v_{\text{diag}}^3 + \frac{v_{\text{diag}}^6}{2 K_{\text{diag}}^2} \right)
\nonumber \\
&+ \frac{1}{2}\left( \frac{a}{F_a} \right)^2 
6 \left( \frac{ v_{\text{diag}}^6}{2 K_{\text{diag}}^2} + \frac{ v_{\text{diag}}^3 v^9}{4 K^8} + \frac{v^6 v_{\text{diag}}^3}{2 K^8} m_u \Lambda_u^2 + \frac{v^6 v_{\text{diag}}^3}{2 K^8} m_d \Lambda_d^2 \right)
\nonumber \\
&+ \frac{1}{2} \left( \frac{\eta'}{F_{\eta'}} \right)^2
\left( m_u v^3 + m_d v^3+\frac{ v_{\text{diag}}^3 v^9}{K^8}+ \frac{v^6 v_{\text{diag}}^3}{2 K^8} m_u \Lambda_u^2 + \frac{v^6 v_{\text{diag}}^3}{2 K^8} m_d \Lambda_d^2 \right)
\nonumber \\
&+ \frac{1}{2} \left( \frac{\pi^0}{F_{\pi^0}} \right)^2 
\left( m_u v^3 + m_d v^3+ \frac{v^6 v_{\text{diag}}^3}{2 K^8} m_u \Lambda_u^2 + \frac{v^6 v_{\text{diag}}^3}{2 K^8} m_d \Lambda_d^2 \right) \nonumber \\
&+ \left( \frac{\eta_d}{F_a} \right) \left( \frac{a}{F_a} \right)
\left( 2\sqrt{6} \frac{ v_{\text{diag}}^6}{2 K_{\text{diag}}^2} \right)\nonumber \\
&+ \left( \frac{a}{F_a} \right) \left( \frac{\eta'}{F_{\eta'}} \right) \left( \sqrt{6} \right) \left( \frac{v_{\text{diag}}^3 v^9}{2 K^8} + \frac{v^6 v_{\text{diag}}^3}{2 K^8} m_u \Lambda_u^2 + \frac{v^6 v_{\text{diag}}^3}{2 K^8} m_d \Lambda_d^2 \right)
\nonumber \\
&+ \left( \frac{\eta'}{F_{\eta'}} \right) \left( \frac{ \pi^0}{F_{\pi^0}} \right) \left( m_u v^3 - m_d v^3 \right) \ , 
\label{eq:axionPotentialGaillardExpansion}
\end{align}
where we have again used the approximation $m_u \Lambda_u^2 - m_d \Lambda_d^2 \approx 0$. We also introduce a compact notation,
\begin{align}
m_+ &= m_u + m_d \ , \quad m_- = m_d - m_u \ , \quad \mu = \frac{m_u m_d}{m_u + m_d} \ , 
\nonumber \\
\mu L^2 &= m_u \Lambda_u^2 + m_d \Lambda_d^2 \ , \quad \Lambda_{\text{inst}}^3 = \frac{L^2}{4K^8} v^6 v_{\text{diag}}^3 \ , \quad \Lambda_{\eta'}^4 = \frac{ v_{\text{diag}}^3 v^9}{4 K^8} \ ,
\nonumber \\
\Lambda_{\text{SSI}}^4 &= K'v_{\text{diag}}^3 \ ,\quad \Lambda_{\text{diag}}^4=\frac{v_{\text{diag}}^6}{2 K_{\text{diag}}^2} \ .
\label{eq:GaiNewVariables}
\end{align} 
The symmetric PNGB mass matrix is now
\begin{align}
\mathcal{L} &=
\frac{1}{2} \left( \eta_d \quad a \quad \eta' \quad \pi^0 \right)
M_{\text{M1}}^2 \left(  \eta_d \quad a \quad \eta' \quad \pi^0 \right)^T \ , 
\label{eq:massMatrixGaillard}
\end{align}
with entries
\begin{align}
\left( M_{\text{M1}}^2 \right)_{11} & = \frac{4}{F_a^2}( \Lambda_{\text{SSI}}^4 + \Lambda_{\text{diag}}^4) \ , &&\left( M_{\text{M1}}^2 \right)_{12} = \frac{1}{F_a^2} 2 \sqrt{6} \Lambda_{\text{diag}}^4 \ ,  \nonumber \\
\left( M_{\text{M1}}^2 \right)_{22} & = \frac{6}{F_a^2} (\Lambda_{\text{diag}}^4 + \Lambda_{\eta'}^4 + 2 \mu \Lambda_{\text{inst}}^3 ) \ , &&\left( M_{\text{M1}}^2 \right)_{23} = \frac{2 \sqrt{6}}{F_aF_{\eta'}} ( \Lambda_{\eta'}^4 + \mu \Lambda_{\text{inst}}^3 )  \ ,  \nonumber \\
\left( M_{\text{M1}}^2 \right)_{33} & = \frac{1}{F_{\eta'}^2}( m_+ v^3 + 4 \Lambda_{\eta'}^4 + 2\mu \Lambda_{\text{inst}}^3)\ , &&\left( M_{\text{M1}}^2 \right)_{34} = \frac{-1}{F_{\pi^0} F_{\eta'}} m_- v^3 \ , \nonumber \\
\left( M_{\text{M1}}^2 \right)_{44} & = \frac{1}{F_{\pi^0}^2} ( m_+ v^3 + 2\mu  \Lambda_{\text{inst}}^3) \ , &&
\left( M_{\text{M1}}^2 \right)_{13} = \left( M_{\text{M1}}^2 \right)_{14} = \left( M_{\text{M1}}^2 \right)_{24} = 0 \ .
\label{eq:massMatrixGaillardEntries}
\end{align}
This result agrees with Ref.~\cite{Gaillard:2018xgk}, validating our derivation procedure.  Again, we diagonalize a generalized form of this mass matrix in the Appendices~\ref{subsec:genMatrix4x4} and~\ref{subsec:genMatrix4x4Expansions}. We use the similarity of the lower right part of the mass matrix to the previous mass matrix in~\eqnref{massMatrixentries} to simplify the eigenvalues of the new axion and the $\eta_d$. The eigenvalues of the $\eta'$ and the $\pi^0$ remain the same as the standard axion story in~\eqnref{pionEtaMass3x3} up to corrections of order $\order{\Lambda_\text{QCD} / \Lambda_{\text{diag}}}$ or $\order{\Lambda_\text{QCD} / \Lambda_{\text{SSI}}}$. For the axion and $\eta_d$, we get
\begin{align}
m_{a}^2 F_a^2 &= 4 (\Lambda_{\text{diag}}^4 + \Lambda_{\text{SSI}}^4) - 24 \Lambda_{\text{diag}}^8 
\nonumber \\
&\times 
\abs{ 2 \Lambda _{\text{SSI}}^4 - \Lambda_{\text{diag}}^4 - 3 (m_a^2 F_a^2)^\text{KSVZ}  - \sqrt{ \left(2 \Lambda _{\text{SSI}}^4 -  \Lambda_{\text{diag}}^4 - 3 (m_a^2 F_a^2)^\text{KSVZ} \right)^2 + 24 \Lambda_{\text{diag}}^8 } }^{-1} \label{eq:Gaillardaxionmass} \ , \\
m_{\eta_d}^2 F_a^2 &= 2 \Lambda _{\text{SSI}}^4 + 5 \Lambda_{\text{diag}}^4 + 3 (m_a^2 F_a^2 )^\text{KSVZ} 
+ \sqrt{ \left(2 \Lambda _{\text{SSI}}^4 - \Lambda_{\text{diag}}^4 - 3 (m_a^2 F_a^2)^\text{KSVZ}\right)^2 + 24 \Lambda_{\text{diag}}^8 } \label{eq:GaillardEtaDmass} \ ,
\end{align}
where $(m_a^2 F_a^2)^\text{KSVZ}$ is defined in~\eqnref{KSVZaxionMassEqu}.  Our results are more precise than those in Ref.~\cite{Gaillard:2018xgk} since we include the mass mixings with the SM mesons.  Their results can be recovered in the limit $\Lambda_{\text{diag}} \ll \Lambda_{\text{SSI}}$ to obtain $m_a^2 F_a^2 \approx 6 \Lambda_{\text{diag}}^4$ and $m_{\eta_d}^2 F_a^2 \approx 4 \Lambda_{\text{SSI}}^4$.

Interestingly, our result also encodes other benchmark axion models, including a simple composite axion~\cite{Kim:1984pt} as well as the KSVZ axion~\cite{Kim:1979if, Shifman:1979if}.  Namely, for the simple composite axion model, we reproduce the hidden axieta and invisible axion mass relations by discarding $\Lambda_{\text{SSI}}$ and expanding in $(m_a^2 F_a^2)^{\text{KSVZ}} / \Lambda_{\text{diag}}^4$.

For the KSVZ axion, we can simply take the limit $\Lambda_{\text{diag}}$, $\Lambda_{\text{SSI}} \to 0$, which recovers the KSVZ axion mass relationship up to an additional group factor arising from the $Q$ field.

\begin{figure}[tb!]
\centering
\begin{subfigure}{0.49\textwidth}
\centering
\includegraphics[width=\textwidth]{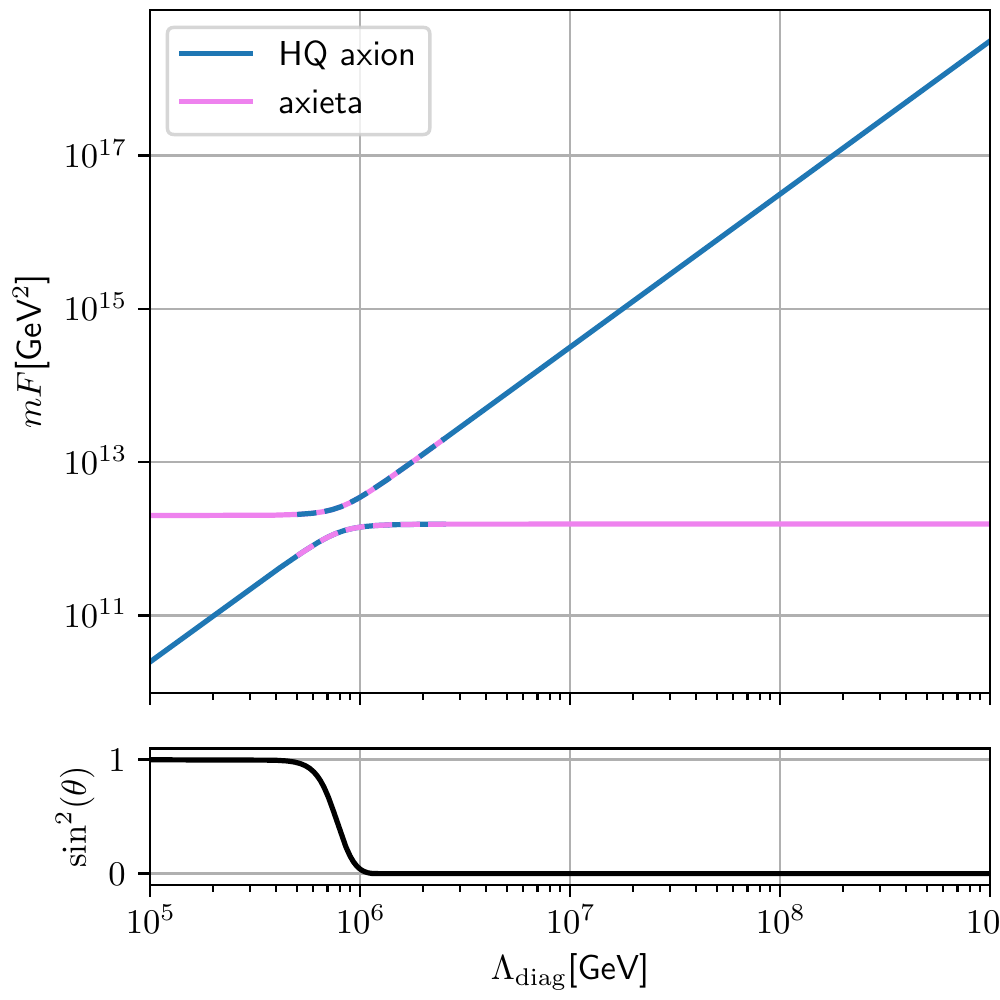}
\subcaption{\label{subfig:massD}}
\end{subfigure}
\hfill
\begin{subfigure}{0.49\textwidth}
\centering
\includegraphics[width=\textwidth]{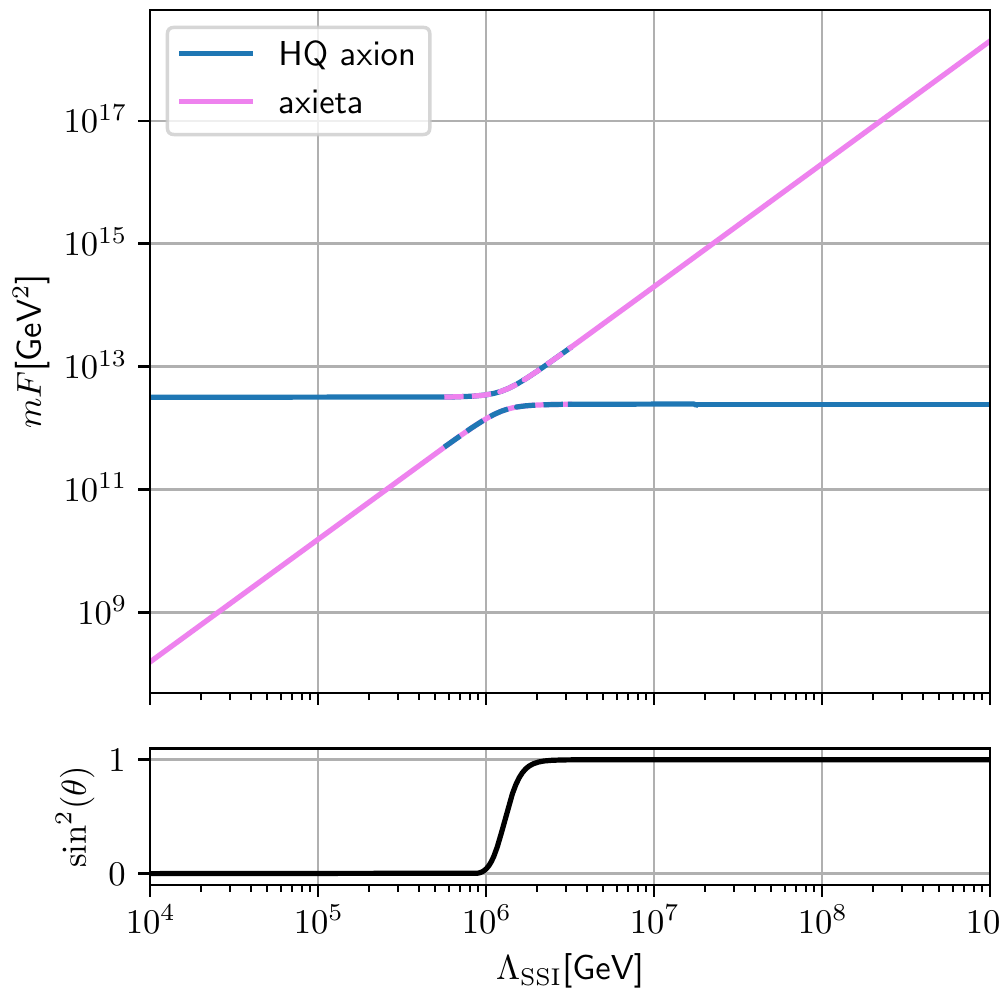}
\subcaption{\label{subfig:massSSI}}
\end{subfigure}
\caption{The axion and axieta masses in~\eqnsref{Gaillardaxionmass}{GaillardEtaDmass} as a function of~(\subref{subfig:massD}) $\Lambda_{\text{diag}}$ with $\Lambda_{\text{SSI}}=10^6$~\text{GeV} and~(\subref{subfig:massSSI}) $\Lambda_{\text{SSI}}$ with  $\Lambda_{\text{diag}}=10^6$~\text{GeV}.  The avoided crossing behavior is highlighted at the non-degenerate symmetry point with a dashed line. In the lower panels, we show the mixing angle for the axion component of the heavier mass eigenstate. 
\label{fig:massesPlot}}
\end{figure}

We present the axion and axieta masses as a function of $\Lambda_{\text{diag}}$ and also $\Lambda_{\text{SSI}}$ in~\figref{massesPlot}, showing how the two eigenvalues exhibit an avoided crossing behavior intrinsic to the mass matrix structure.  Since both particles are PNGBs of the same symmetry, they are indistinguishable at the symmetric point for $\Lambda_{\text{diag}} = \Lambda_{\text{SSI}}$ and only a small splitting arises from QCD effects.  In addition, the ``gauge" eigenstate of the mass eigenvalues is flipped as we change the hierarchy between $\Lambda_{\text{diag}}$ and $\Lambda_{\text{SSI}}$, shown via the mixing angles in the lower panels of~\figref{massesPlot}.


\subsubsection{Axion and axieta masses in the M2 variant of the color unification model}

We now discuss model M2, defined in Table~\ref{subtab:fieldContentM2}, which includes a second scalar field $\Delta_2$ instead of the exotic quark $q$. As explained in Ref.~\cite{Gaillard:2018xgk}, the second scalar field carries PQ-charge and has a vev $\langle \Delta_2 \rangle =\Lambda_\text{CUT}$ which sets $F_a = \Lambda_\text{CUT}$. Compared to the first model M1, where the PQ breaking scale was $\Lambda_\text{diag}$, we now have two axion-like degrees of freedom $a'$ and $a$ with different PQ scales $F_a$ and $F_d$, respectively.

Most of the steps from~\eqnref{BigQcondensate} to~\eqnref{massMatrixGaillardEntries} remain the same and we briefly comment on the differences.  First, the exotic quark condensate is only composed of $\langle \bar{Q} Q\rangle$ as in~\eqnref{BigQcondensate}. 
Second, for the axion potential we consider the 't Hooft determinental operator in~\eqnref{determInteraction}, where we replace the phase $\exp(-ic_3^G \frac{a}{F_a})$ by $\exp(-i c_3^{G'} \frac{a'}{F_a})$ to account for the new axion $a'$, where $c_3^{G'} = 12$ from Ref.~\cite{Gaillard:2018xgk}. Finally, the first two instanton diagrams in Fig.~\ref{subfig:fig4_bubble1} and Fig.~\ref{subfig:fig4_bubble2} change since there is a heavy exotic quark mass insertion instead of the quark condensate $\bar{q}q$, and the new instanton amplitudes are now
\begin{align}
    A_1 & = 
    2 m_{q'}\Lambda_{q'}^2 \cos(12 \frac{a'}{F_a}) \ , \\
    A_2 & = 
    m_{q'}\Lambda_{q'}^2 v_\text{diag}^3 \cos(\sqrt{6} \frac{a}{F_d} - 12 \frac{a'}{F_a}) \ .
\end{align}

We again introduce a compact notation for the new scales replacing
\begin{align}
\Lambda_{\text{SSI}}^4 = 2 K' m_{q'}\Lambda_{q'}^2 \ , \text{ and } \quad \Lambda_{\text{diag}}^4 = \frac{m_{q'}\Lambda_{q'}^2 v_\text{diag}^3}{K_\text{diag}^2} \ ,
\end{align}
where other previous definitions from~\eqnref{GaiNewVariables} remain. Our PNGB mass matrix becomes
\begin{align}
\mathcal{L} &=
\frac{1}{2} \left( a' \quad a \quad \eta' \quad \pi^0 \right)
M_{\text{M2}}^2 \left(  a' \quad a \quad \eta' \quad \pi^0 \right)^T \ , 
\label{eq:massMatrix2Gaillard} 
\end{align}
with new entries in the upper left part given by
\begin{align}
\left( M_{\text{M2}}^2 \right)_{11} & = \frac{144}{F_a^2}( \Lambda_{\text{SSI}}^4 + \Lambda_{\text{diag}}^4) \ , &&\left( M_{\text{M2}}^2 \right)_{12} = \frac{-12 \sqrt{6}}{F_a F_d}  \Lambda_{\text{diag}}^4 \ ,  \nonumber \\
\left( M_{\text{M2}}^2 \right)_{22} & = \frac{6}{F_d^2} (\Lambda_{\text{diag}}^4+\Lambda_{\eta'}^4 + 2 \mu \Lambda_{\text{inst}}^3) &&\left( M_{\text{M2}}^2 \right)_{23} = \frac{2 \sqrt{6}}{F_d F_{\eta'}} ( \Lambda_{\eta'}^4 + \mu \Lambda_{\text{inst}}^3 ) \label{eq:massMatrix2GaillardEntries}\ ,
\end{align}
where the remaining entries match $M_\text{M1}^2$ in~\eqnref{massMatrixGaillardEntries}. The result in~\eqnref{massMatrix2GaillardEntries} agrees with the mass matrix for M2 in Ref.~\cite{Gaillard:2018xgk}, again validating our instanton amplitudes calculation framework.  The new mass eigenstates for $a'$ and $a$, corresponding to~\eqnsref{Gaillard2AxionPrimeMassApp}{Gaillard2AxionMassApp}, are
\begin{align}
m_{a'}^2 F_a^2 &= 144 \left(\Lambda_{\text{diag}}^4 + \Lambda _{\text{SSI}}^4 - \Lambda_{\text{diag}}^4 \left(1 + \frac{(m_a^2 F_a^2)^\text{KSVZ}}{\Lambda_{\text{diag}}^4}\right)^{-1} \right)  \ , \label{eq:Gaillard2AxionPrimeMass} \\
m_{a}^2 F_d^2 &= 6 \Lambda_{\text{diag}}^4 + 6 (m_a^2 F_a^2)^\text{KSVZ} + 144 \frac{F_d^2}{F_a^2} \left( \Lambda_{\text{diag}}^4 \left( 1 + \frac{(m_a^2 F_a^2)^\text{KSVZ}}{\Lambda_{\text{diag}}^4} \right)^{-1} \right)\ . \label{eq:Gaillard2AxionMass}
\end{align}
In contrast to model M1, the eigenvalues do not show an avoided crossing behavior, since model M2 includes a scale separation given by $F_a \sim \Lambda_\text{CUT}\gg F_d$. While our result includes the mixing effects between $a'$ and $a$ with SM mesons, we can reproduce the simplified result in Ref.~\cite{Gaillard:2018xgk} by taking the limit $(m_a^2 F_a^2)^\text{KSVZ}\ll \Lambda_{\text{diag}}^4\ll \Lambda _{\text{SSI}}^4$ to obtain
\begin{align}
m_{a'}^2 F_a^2 & \approx 144 \Lambda _{\text{SSI}}^4 \ , \qquad m_{a}^2 F_d^2 \approx 6 \Lambda_{\text{diag}}^4 \ .
\end{align}
Having derived the mass eigenstates for the axions in the two model variants M1 and M2, we can now calculate the corresponding electromagnetic coupling $G_{a \gamma \gamma}$ precisely, since our framework includes the mixing effects with SM mesons.





\subsection{The axion-diphoton coupling with SSI effects}

First, we need to generalize the diphoton coupling in~\eqnref{elMagLagrEigensystem} to the case of two axions. In this case we have a $4\times 4$ eigensystem that contains four PNGB mass eigenstates defined by
\begin{align}
    \mqty(a_{1,m}\\a_{2,m}\\\eta'_m\\\pi^0_m) = V^T \mqty(a_1\\a_2\\\eta'\\\pi^0),
\end{align}
where $V$ is defined as $(\vec{v}_1 | \vec{v}_2 | \vec{v}_3 | \vec{v}_4)$ for normalized eigenvectors $\vec{v}_1,\vec{v}_2,\vec{v}_3$ and $\vec{v}_4$. The states $a_1$ and $a_2$ are later attributed to the appropriate axion-like states of M1 or M2 model variants. We obtain the net electromagnetic coupling for the axion mass eigenstates as a coherent sum from the corresponding ``gauge" eigenstate PNGBs, weighted by the eigenvector entries that are approximated in a $\mathcal{O}(v / F_a)$ expansion in case of $a_1$ or $\mathcal{O}(v / F_d)$ expansion in case of $a_2$.  Correspondingly, the mass basis axion interactions with photons is
\begin{align}
\nonumber
\mathcal{L} &\supset  -\frac{1}{4}\left( \left(\frac{\alpha_e}{2\pi F_a} E_1 \right)v_{1,1} + \left(\frac{\alpha_e}{2\pi F_d} E_2 \right)v_{1,2} + G_{\eta'\gamma\gamma} v_{1,3} + G_{\pi^0\gamma\gamma} v_{1,4}\right) a_{1,m} F_{\mu\nu} \tilde{F}^{\mu\nu}\\
&\quad -\frac{1}{4}\left(  \left(\frac{\alpha_e}{2\pi F_a} E_1 \right)v_{2,1} + \left(\frac{\alpha_e}{2\pi F_d} E_2 \right)v_{2,2} + G_{\eta'\gamma\gamma} v_{2,3} + G_{\pi^0\gamma\gamma} v_{2,4}\right) a_{2,m} F_{\mu\nu} \tilde{F}^{\mu\nu}
\nonumber \\
&\simeq -\frac{1}{4} \left( \frac{\alpha_e}{2\pi F_a} \right) \left( E_1  - \Delta_1 \right) a_{1,m} F_{\mu\nu} \tilde{F}^{\mu\nu} - \frac{1}{4} \left( \frac{\alpha_e}{2\pi F_d} \right) \left( \frac{F_d}{F_a} E_1 v_{2,1} - \Delta_2 \right) a_{2,m} F_{\mu\nu} \tilde{F}^{\mu\nu} \ , \label{eq:defDiphotonCouplinGai}
\end{align}
with
\begin{align}
\Delta_1 &\equiv -\frac{2\pi}{\alpha_e}\left( G_{\eta'\gamma\gamma}(F_a v_{1,3})+ G_{\pi^0\gamma\gamma}(F_a v_{1,4}) \right) \ , \\
\Delta_2 &\equiv -\frac{2\pi}{\alpha_e}\left( G_{\eta'\gamma\gamma}(F_d v_{2,3})+ G_{\pi^0\gamma\gamma}(F_d v_{2,4}) \right) \ ,
\label{eq:defDeltasGaillard}
\end{align}
such that $\Delta_1$ and $\Delta_2$ encode the correction arising from mixing effects and $E_1, E_2$ describe the contact coupling of photons to the corresponding unmixed axions. In~\eqnref{defDiphotonCouplinGai} we have used the fact that $E_2 = 0$ in both model variants.


In context of the first model M1, we identify in~\eqnref{defDiphotonCouplinGai} $a_1 = a, a_2 = \eta_d$ and $F_d = F_a$. Since the PQ-charged exotic quarks $Q$ and $q$ have no electroweak quantum numbers, the contact couplings $E_1$ and $E_2$ vanish trivially. For the mass matrix in~\eqnref{massMatrixGaillardEntries} the necessary eigenvector components are based on~\eqnsref{eVec14x4M1}{eVec24x4M1} in the Appendix using the mass matrix entries in~\eqnref{massMatrixGaillardEntries}, with non-trivial dependence on the scales $\Lambda_{\text{SSI}}$ and $\Lambda_{\text{diag}}$.
Note that the diphoton coupling of M1 depends non-trivially on the scales $\Lambda_{\text{SSI}}$ and $\Lambda_{\text{diag}}$.  Our result is valid in different hierarchies of $\Lambda_{\text{SSI}}$ and $\Lambda_{\text{diag}}$ or, equivalently, for different instanton amplitudes $K', K_{\text{diag}}$ and $K$ of the gauge groups $SU(6), SU(3)_{\text{diag}}$ and $SU(3)_c$, respectively. Ref.~\cite{Gaillard:2018xgk} relates the scales $\Lambda_{\text{CUT}}, \Lambda_{\text{SSI}}$ and $\Lambda_{\text{diag}}$ to each other by the approximation
\begin{align}
    \left( \frac{\Lambda_{\text{SSI}}}{\Lambda_{\text{diag}}} \right)^4 \simeq 4.5\cdot 10^{-10} \left( \frac{\Lambda_{\text{CUT}}}{\Lambda_{\text{diag}}} \right) \ , \label{eq:scaleRelationGai}
\end{align}
where the running coupling constant of $SU(3')$ was set at the scale $\Lambda_{\text{CUT}} = 10^{14}$~GeV to be $\alpha'(\Lambda_{\text{CUT}}) = 0.3$. 

\begin{figure}[htb]
\centering
\includegraphics[width=\textwidth]{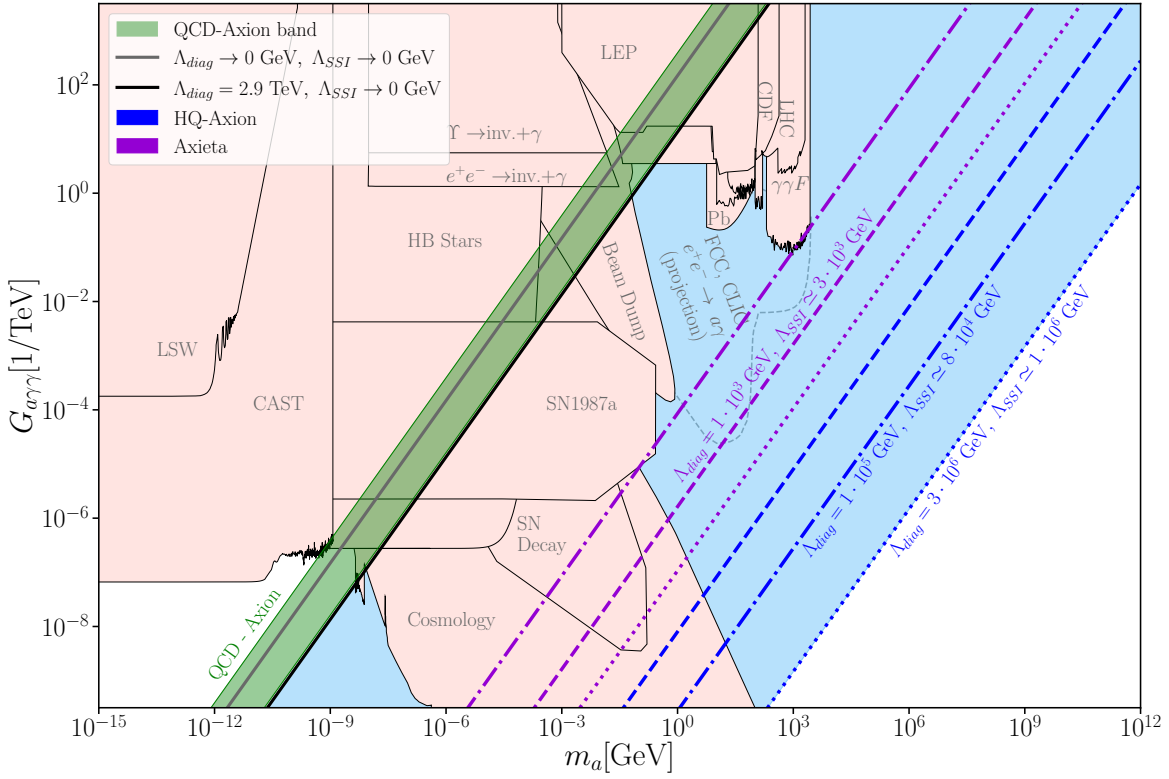}
\caption[Resulting axion bands in the ($m_a$, $G_{a\gamma\gamma}$)-parameter space]
{Axion bands of the color unified model variant M1 in the $m_a$ vs.~$G_{a\gamma\gamma}$ plane. The band in green shows canonical DFSZ- and KSVZ-models, while the pink shaded regions show the current experimental bounds on axions and ALPs~\cite{Jaeckel:2010ni,Alekhin:2015byh,Jaeckel:2015jla,Redondo:2008en,Cadamuro:2011fd,Proceedings:2012ulb,Jaeckel:2012yz,Mimasu:2014nea,Payez:2014xsa,Millea:2015qra,Jaeckel:2017tud,CAST:2017uph}, and we include a contour from a future projection of CLIC sensitivity from Ref.~\cite{Bauer:2017ris}.  The blue shaded region shows the possible parameter space for the new HQ-axion $a$ and the axieta $\eta_d$ enhanced by SSI effects, where lines in blue and violet show chosen models with the corresponding scales $\Lambda_{\text{diag}}$ and $\Lambda_{\text{SSI}}$. A lower and upper bound on the $\Lambda_{\text{diag}}$ and $\Lambda_{\text{SSI}}$ scales arises because of experimental constraints on new color gauge group extensions discussed in the main text.}
\label{fig:gaiFaCorrM1}
\end{figure}

We show the diphoton coupling vs.~SSI enhanced axion mass in~\figref{gaiFaCorrM1} for model variant M1.

We plot representative choices of the two underlying scales for the extended color symmetry  $(\Lambda_{\text{diag}}$, and the SSI scale $\Lambda_{\text{SSI}})$, where each choice gives rise to two axion degrees of freedom, the high-quality axion $a$ and the exotic axieta $\eta_d$.  The indicated parameter space of all viable choices $(\Lambda_{\text{diag}}$, $\Lambda_{\text{SSI}})$ is shaded in blue and is mostly outside of the traditional QCD axion band, which is shaded in green.  In particular,
the null result for the color-octet pion $\pi_d$ from LHC experimental searches for scalar gluon resonances constrains $\Lambda_{\text{diag}} > 2.9$~TeV~\cite{Gaillard:2018xgk}.  This restricts the blue region to the right of the solid black line and also coincidentally delineates the QCD axion band from the new axion parameter space.  For reference, we also show that our axions lie within the QCD axion band when we eliminate both the exotic color gauge group and the SSI effects, indicated by the dark gray line in~\figref{gaiFaCorrM1}.

\begin{figure}[htb]
\centering
\includegraphics[width=\textwidth]{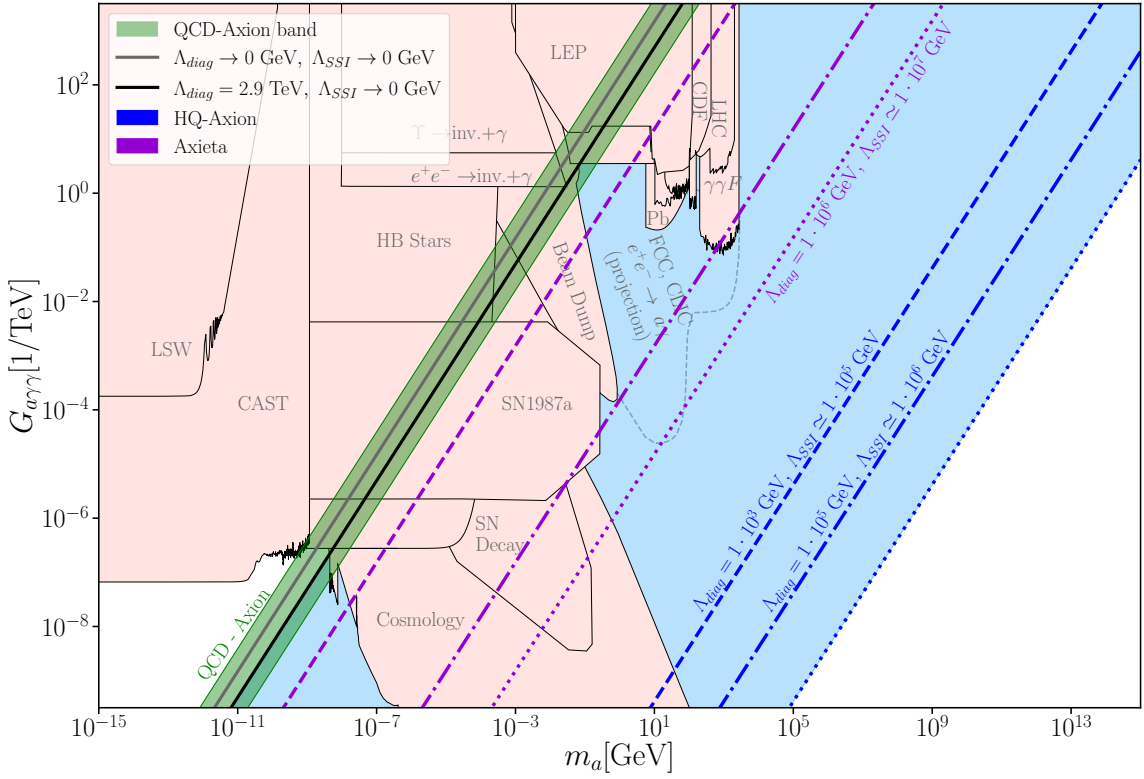}
\caption[Resulting axion bands in the ($m_a,G_{a\gamma\gamma}$)-parameter space]
{Axion bands of the color unified model variant M2 in the $m_a$ vs.~$G_{a\gamma\gamma}$ plane. The band in green shows canonical DFSZ- and KSVZ-models, while the pink shaded regions show the current experimental bounds on axions and ALPs~\cite{Jaeckel:2010ni,Alekhin:2015byh,Jaeckel:2015jla,Redondo:2008en,Cadamuro:2011fd,Proceedings:2012ulb,Jaeckel:2012yz,Mimasu:2014nea,Payez:2014xsa,Millea:2015qra,Jaeckel:2017tud,CAST:2017uph}, and we include a contour from a future projection of CLIC sensitivity from Ref.~\cite{Bauer:2017ris}.  The blue shaded region shows the possible parameter space for the new HQ-axion $a'$ and the axieta $a$ enhanced by SSI effects, where lines in blue and violet show chosen models with the corresponding scales $\Lambda_{\text{diag}}$ and $\Lambda_{\text{SSI}}$.  A lower and upper bound on the $\Lambda_{\text{diag}}$ and $\Lambda_{\text{SSI}}$ scales arises because of experimental constraints on new color gauge group extensions discussed in the main text.}
\label{fig:gaiFaCorrM2}
\end{figure}

In model variant M2, we identify $a_1 = a'$ and $a_2 = a$. In contrast to M1, we have a contact diphoton coupling for $a'$ induced by nine exotic quarks with PQ-charge $+1$, which share the same hypercharge with the SM chiral quark fields.  The contact couplings are therefore readily calculated according to Ref.~\cite{DiLuzio:2020wdo} to be $E_1 = 5$ and trivially $E_2 = 0$. We recall that there is no bifundamental fermion with $SU(3')$ and $SU(3)_\text{diag}$ color charge and thus the instanton scale $\Lambda_{\text{SSI}}$ is not related to the confinement scale of $SU(3)_{\text{diag}}$. 
Thus, in contrast to M1, the choices of $\Lambda_{\text{diag}}$ and $\Lambda_{\text{SSI}}$ are independent. Also, the hierarchy $F_d\ll F_a$ implies that we can neglect the $v_{2,1}$ term in~\eqnref{defDiphotonCouplinGai} . Our results for $a_1$ and $a_2$ masses and diphoton couplings are shown in~\figref{gaiFaCorrM2}.

Recently, SSI effects were shown to affect EDM searches~\cite{Bedi:2022qrd}, since the SSI effects can push the axion potential away from the energetic minimum and manifest as an observable nEDM.  In our case, Ref.~\cite{Bedi:2022qrd} sets an upper bound  $\Lambda_{\text{SSI}} \lesssim 10^{-8} \Lambda_{\text{CUT}}$ and $\Lambda_{\text{SSI}} \lesssim 10^{-7} \Lambda_{\text{CUT}}$ for M1 and M2 variants, respectively, which restricts the blue regions from the right in~\figsref{gaiFaCorrM1}{gaiFaCorrM2}.


We conclude that composite axion models featuring an extension of the color gauge symmetry generally lie outside of the traditional QCD axion band due to exotic mass mixing effects and possible SSI effects. Such axions are naturally within the realm of high-energy collider experiments and can feature relatively large diphoton couplings, a standard but critical signature at colliders.  The collider phenomenology of heavy pseudoscalar particles has also been revisited recently in the context of axion-like particles (see, e.g.~\cite{Bauer:2017ris}), but we emphasize that the new, heavy degree of freedom in such searches can be a true QCD axion responsible for solving the strong CP problem.

In summary, we have derived the diphoton coupling of the high quality QCD-axion influenced by SSI effects and we have shown the present constraints in~\figsref{gaiFaCorrM1}{gaiFaCorrM2} for two exemplary models based on Ref.~\cite{Gaillard:2018xgk}. The successful application of our framework allows for a systematic consideration of all given constraints for axion-like particles in context of HQ and composite axions. In addition to diphoton constraints from axion searches, the parameter space of composite axion models includes bounds for $\Lambda_{\text{diag}}$ and $\Lambda_{\text{SSI}}$ from above and below, where the bound from below is driven by collider constraints searching for exotic colored states and the bound from above comes from possible regeneration of an observable neutron EDM. Although both of these probes will improve in future experiments, the complementary direct searches for high quality axions in the reach of colliders will be critical in exploring the new parameter space of SSI-enhanced axion models.





\section{Conclusion}
\label{sec:conclusion}

In this work we have studied the small size instanton effects arising from confinement of high scale $SU(N)$ symmetries that embed our color gauge symmetry. 
While standard QCD axion calculations are insufficient for including SSI effects, we demonstrate that our approach based on 't Hooft determinantal interaction with instanton amplitudes consistently includes the masses of low-energy PNGBs as well as mixing effects in the PNGB diphoton couplings. Based on the color unified theory of Ref.~\cite{Gaillard:2018xgk}, we extended the color gauge group to a product $SU(6) \times SU(3')$ symmetry, leading to two axion particles that have SSI-enhanced  properties.  Our main results are summarized in~\figsref{gaiFaCorrM1}{gaiFaCorrM2}, which show that SSI effects dramatically expand the region of QCD axion models into the $m_a$ vs.~$G_{a \gamma \gamma}$ plane.  In particular, the resulting axions can have masses at the electroweak scale or even heavier, leading to resonance signatures at high energy colliders.


Hence, searches for high mass QCD axions at colliders are well justified. We emphasize that in these models an extension of the color symmetry is necessary, which ties together the search for new exotic colored states and axions via their effective operators. Moreover the new exotic colored states will generically have a new exotic axieta particle which plays a role in solving the strong CP problem.  The simultaneous discovery of two distinct axion particles along with a family of new exotic colored states at high energy colliders would be a striking signature of these axion models that feature SSI effects. 

\section*{Acknowledgments}
\label{sec:acknowledgments}

This research is supported by the Cluster of Excellence PRISMA$^+$, ``Precision Physics, Fundamental Interactions and Structure of Matter" (EXC 2118/1) within the German Excellence Strategy (project ID 39083149).  FY would like to express special thanks to the Mainz Institute for Theoretical Physics (MITP) of the Cluster of Excellence PRISMA+ (Project ID 39083149), for its hospitality and support.
\begin{appendix}

\input{appendix}

\end{appendix}

\bibliographystyle{apsrev4-1}
\bibliography{quotations,cajohareRefs}

\end{document}

%% file: appendix.tex
\section{Topological susceptibility}
\label{sec:topSusc}

This section aims to give a brief review of the topological susceptibility of QCD and its connection to instantons, following Refs.~\cite{Leutwyler:1992yt, Huang:1993cf}. In~ \secref{instantons}, we discussed how the background instanton field affects the colored fermions via the effective 't Hooft determinantal operator.  In essence, the operator accounts for quantum effects in the QCD Lagrangian where gluons, including instanton configurations, are integrated out.  

Introducing the determinantal interaction is necessary to account for the instanton induced potential and for the mass mixings between all pseudoscalars in the low energy spectrum.  To see this, we need to consider QCD below the confining scale, which necessitates integrating out the $\theta$-term, $\theta G \tilde{G}$.  Since the $\theta$-term is modified by the quark $U(1)$ axial anomaly, the CP-violating amplitude of QCD scales with the parameter $\bar{\theta}$.  We can thus study CP-violating effects in low-energy QCD by performing at a moment expansion of the vacuum-to-vacuum amplitude in $\bar{\theta}$, where the topological susceptibility $\mathcal{T}$ is defined as the second moment.  Hence, the topological susceptibility is a quantity that gives a measure of instanton effects on the pseudoscalar masses.

We derive the topological susceptibility by first considering the generating functional for QCD in Euclidean space above the confinement scale,
\begin{align}
Z(\bar{\theta}) = \sum_{\nu\in\mathbb{Z}}\int [DG_\mu][D q][D\bar{q}] e^{i\bar{\theta}\nu} \exp 
\left[ -\int  \dd^4 x_E \left( \sum_{i = u,d,s} \bar{q}_i (\slashed{D}+m_{i}) q_i+\frac{1}{4} G_{\mu \nu} G^{\mu \nu} \right) \right] \ ,
\end{align}
where we discard gauge fixing and ghost terms for simplicity and the quark masses are diagonal and real, allowing us to write the winding number phase as $e^{i \bar{\theta} \nu}$~\cite{Huang:1993cf}.  Note that we have not integrated out instanton field configurations yet, which is why the sum over winding numbers $\nu$ is included in the path integral. To perform this step, we make use of the dilute gas approximation (DGA), which sums over all instanton field configurations with different winding numbers, originated in Refs.~\cite{tHooft:1976rip, tHooft:1986ooh}.  The result is
\begin{align}
Z(\bar{\theta}) = \exp(2V_E K \cos(\bar{\theta}) m_u m_d m_s) \ ,
\end{align}
where $V_E$ is the four-space Euclidean volume and $K$ is the instanton amplitude~\cite{Callan:1976je, Callan:1977gz, Huang:1993cf}. 

We use this result to calculate the first and second moment of $\log(Z(\bar{\theta}))$. The first moment becomes the expectation value $\langle G\tilde{G}\rangle$ and the second moment defines the topological susceptibility $\mathcal{T}$. Dividing by the four-space volume $V_E$, we get
\begin{align*}
\langle G\tilde{G}\rangle &= \frac{1}{V_E} \left\langle \int \dd^4 x~i G\tilde{G}\right\rangle = \frac{1}{V_E} \dv{}{\bar{\theta}} \log(Z(\bar{\theta})) = -2 m_u m_d m_s K\sin(\bar{\theta}) \ , \\[10pt]
\mathcal{T}&=\frac{1}{V_E} \dv[2]{}{\bar{\theta}}\log(Z(\bar{\theta})) = -2 m_u m_d m_s K\cos(\bar{\theta}) \ .
\end{align*}
For very small $\bar{\theta}$ we can relate both values to each other:
\begin{align*}
\langle G\tilde{G}\rangle \approx -2 m_u m_d m_s K \bar{\theta} \approx \mathcal{T}\bar{\theta} \ .
\end{align*} 
The topological susceptibility for QCD is calculated to be of order $\mathcal{T} \sim m_\pi^2 f_{\pi}^2$ using current algebra~\cite{Shifman:1979if}, and we coincidentally observe $\mathcal{T} \sim m_\pi^2 f_{\pi}^2 \sim \Lambda_{\text{QCD}}^4$ numerically~\cite{Shifman:2012zz}.

Given that the axion field is a dynamical variable for $\bar{\theta}$, we see that $\mathcal{T}$ is the axion mass contribution from low energy QCD, since $\langle G \tilde{G} \rangle$ breaks the axion shift symmetry.  Note that the axion field is the Goldstone boson of a Peccei-Quinn symmetry $U(1)_\text{PQ}$, which is broken only by anomalous effects.  This means that the QCD axion mass $m_a^2 f_a^2$ will be at least $\langle G \tilde{G}\rangle \approx \mathcal{T}$ in any scenario.  In vanilla QCD axion models, $\mathcal{T}$ is solely responsible for the mass of the axion. In~\secref{instantons} we discuss how the instanton amplitude $K$ changes due to SSI effects coming from non-trivial embeddings of $SU(3)_c$.  This also affects the topological susceptibility $\mathcal{T}$ and therefore raises the axion mass.

\section{Mass matrix axion mixing with SM PNGBs}
\label{sec:genMatrix}

In the following, we parametrize a general mass matrix in the case where we have an axion mass, which depends on a scale $F_a$ coming from a $U(1)_{PQ}$ breaking scale. For our solution we deploy a method that is similar to time-independent perturbation theory in quantum mechanics.  Since the necessary solutions have to be up to a certain order in $\frac{1}{F_a}$, we diagonalize the whole eigensystem, the eigenvalues as well as the eigenvectors,  in a power series in $\frac{1}{F_a}$ to calculate the results order by order. This is the same procedure as in perturbation theory  in quantum mechanics, where one aims to obtain the eigenvalues and eigenstates of a power series coming from a small interaction term. We assume that $F_a$ is very large (physically of order $\order{10^{10} \si{GeV}}$) to derive the resulting mass eigenvalues and eigenvectors. Similar to perturbation theory in quantum mechanics, the quality of our result depends on the smallness of the expansion parameter $\frac{1}{F_a}$ and the solution breaks down at order $n \sim F_a$. We present the method in detail for a 3-dimensional mass matrix, and show our results for the 4-dimensional mass matrices.

\subsection{3-dimensional mass matrix}
\label{subsec:genMatrix3x3}
We parametrize the 3-dimensional mass matrix in the following way:
\begin{align}
\nonumber
M^{(3\times3)}&=M_0^{(3\times3)}+\frac{1}{F_a}M_1^{(3\times3)}+\frac{1}{F_a^2}M_2^{(3\times3)}\\
&=\left(
\begin{array}{ccc}
  0& 0& 0 \\
 0 & \beta & \sigma \\
 0 & \sigma & \gamma \\
\end{array}
\right)+\frac{1}{F_a}\left(
\begin{array}{ccc}
 0 & \kappa & \nu \\
 \kappa & 0 & 0 \\
 \nu & 0 & 0 \\
\end{array}
\right)+\frac{1}{F_a^2}\left(
\begin{array}{ccc}
 \alpha & 0 & 0 \\
 0 & 0 & 0 \\
 0 & 0 & 0 \\
\end{array}
\right)
\label{eq:genMassMatrix}
=\left(
\begin{array}{ccc}
 \frac{\alpha}{F_a^2} & \frac{\kappa}{F_a} & \frac{\nu}{F_a} \\
 \frac{\kappa}{F_a} & \beta & \sigma \\
 \frac{\nu}{F_a} & \sigma & \gamma \\
\end{array}
\right) \ .
\end{align}
Similarly, we define the power series for the eigenvectors $\vec{u}$, $\vec{v}$, $\vec{w} $ and eigenvalues $\lambda^{(n)}$ as
\begin{align}
\label{eq:eigEqExpansion}
\vec{u} &\equiv \sum_{i=0}^{\infty}\frac{1}{F_a^i}\vec{u}_i= \vec{u}_0+\frac{1}{F_a}\vec{u}_1+\cdots \ , \qquad
\vec{v} \equiv \sum_{i=0}^{\infty}\frac{1}{F_a^i}\vec{v}_i \ , \qquad 
\vec{w} \equiv \sum_{i=0}^{\infty}\frac{1}{F_a^i}\vec{w}_i \ , \nonumber \\
\lambda^{(n)} &= \sum_{i=0}^{\infty}\frac{1}{F_a^i}\lambda_i^{(n)} = \lambda_0^{(n)}+\frac{1}{F_a}\lambda_1^{(n)} + \ldots \ ,
\end{align}
where $n = 1$, $2$, $3$ corresponds to the eigenvectors $\vec{u}$, $\vec{v}$, $\vec{w}$ respectively. The eigensystem equation can now be expanded order by order in $\dfrac{1}{F_a}$. After reshuffling, we define $\Xi_i$ as the eigenvalue system at order $\dfrac{1}{F_a^i}$, 
\begin{align}
0 &=M\vec{u} - \lambda^{(1)} \vec{u} \nonumber \\
0 &=\left(M_0+\frac{1}{F_a}M_1+\frac{1}{F_a^2}M_2\right) \left(\sum_{i=0}^{\infty}\frac{1}{F_a^i}\vec{u}_i\right) - \left(\sum_{i=0}^{\infty}\frac{1}{F_a^i}\lambda_i^{(1)}\right) \left(\sum_{i=0}^{\infty}\frac{1}{F_a^i}\vec{u}_i\right)  \nonumber \\[5pt]
0 &=\left( M_0 \vec{u_0} - \lambda_0^{(1)} \vec{u_0}\right)+\frac{1}{F_a}\left( M_0 \vec{u_1}+M_1 \vec{u_0} - \lambda_0^{(1)} \vec{u_1}- \lambda_1^{(1)} \vec{u_0}\right) + \ldots
\nonumber \\
0 &= \left( \mqty(0\\ \beta u_0^y+\sigma u_o^z\\ \sigma u_0^y+ \gamma u_o^z)-\lambda_0^{(1)} \mqty(u_0^x\\u_0^y\\u_0^z) \right)+\frac{1}{F_a}\left( \mqty(\kappa u_0^y + \nu u_0^z \\ \kappa u_0^x+\beta u_1^y+\sigma u_1^z\\ \nu u_0^x + \sigma u_1^y+ \gamma u_1^z)-\lambda_0^{(1)} \mqty(u_1^x\\u_1^y\\u_1^z) -\lambda_1^{(1)} \mqty(u_0^x\\u_0^y\\u_0^z) \right)+\ldots \nonumber \\[5pt]
0 &\equiv \sum_{i=0}^{\infty} \frac{1}{F_a^i}\Xi_i^{(1)} ,
\label{eq:orderEigSysEquation}
\end{align}
where the components $u_i^x$, $u_i^y$, $u_i^z$ and $\lambda_i^{(1)}$ can be determined and we can permute for $\vec{v}$ and $\lambda_i^{(2)}$ as well as $\vec{w}$ and $\lambda_i^{(3)}$.  In addition to $\Xi_i^{(1)}$, $\Xi_i^{(2)}$, and $\Xi_i^{(3)}$, we have two additional constraints coming from the orthonormality requirement of the eigenvectors, namely 
\begin{align}
1 &= \vec{u}\cdot \vec{u} = \vec{u}_0 \cdot \vec{u}_0 + \frac{1}{F_a}(2 \vec{u}_1 \cdot \vec{u}_0) + \ldots \equiv \sum_{i=0}^{\infty} \frac{1}{F_a^i}\Delta_i^{(1)} \ , \\
0 &= \vec{u}\cdot \vec{v} = \vec{u}_0 \cdot \vec{v}_0 + \frac{1}{F_a}( \vec{u}_1 \cdot \vec{v}_0 + \vec{u}_0 \cdot \vec{v}_1) + \ldots \equiv \sum_{i=0}^{\infty} \frac{1}{F_a^i}\Theta_i^{(1,2)} \ ,
\end{align}
leading to
\begin{align}
\sum_{i=0}^{\infty} \frac{1}{F_a^i} \Delta_i^{(n)} = 1 \ , \quad \text{ and } \quad \sum_{i=0}^{\infty}
\frac{1}{F_a^i}\Theta_i^{(n,m)} = 0 \text{ for } n,m=1,2,3 \quad, n\neq m \ .
\end{align}
We now take the aforementioned ansatz of perturbation theory to assume that the eigensystems of different orders are independent of each other for large $F_a$:
\begin{align}
\label{eq:orderAnsatz}
\sum_{i=0}^{\infty} \frac{1}{F_a^i}\Xi_i^{(n)} =0 \qquad &\Rightarrow \qquad \Xi_i^{(n)} =0,\qquad \forall i\in \mathbb{N}_0,\quad n\in\{1,2,3\},\\
\sum_{i=0}^{\infty} \frac{1}{F_a^i}\Delta_i^{(n)} = 1  \qquad &\Rightarrow \qquad \Delta_0^{(n)} =1,\quad \Delta_i^{(n)} =0,\qquad \forall i\in \mathbb{N}, \quad n\in\{1,2,3\} ,\\
\sum_{i=0}^{\infty} \frac{1}{F_a^i}\Theta_i^{(n,m)}=0  \qquad &\Rightarrow \qquad \Theta_i^{(n,m)}=0,\qquad \forall i \in \mathbb{N}_0,\quad n,m\in \{1,2,3\}, n\neq m.
\end{align}
Using this ansatz we can solve for the unknowns going order by order upwards. We demonstrate the procedure for $n=1$, which corresponds to the axion mass eigenvalue. At the zeroth order we get, ignoring the trivial solution, three solutions $L_0^n$ that each correspond to one of the three eigenvalues $\lambda^{(n)}$. The corresponding solutions are 
\begin{align}
\Xi_0^{(1)}&=0, \quad\Delta_0^{(1)} =1,\quad \Theta_0^{(1,2)}=0,\Theta_0^{(1,3)}=0 \quad\Rightarrow\quad  L_0^1=\lbrace\lambda_0^{(1)}=0,u_0^x=1,u_0^y=0,u_0^z=0\rbrace\\
\nonumber
\Rightarrow L_0^2= 
& \left\{ \lambda_0^{(2)}=\frac{ (\beta+\gamma)+\sqrt{(\beta-\gamma)^2+4 \sigma^2}}{2},\right. \\
\label{eq:ev2Sol0}
\quad & \hspace{1cm} \left.  v_0=\left( 0,v_0^y,\frac{2\sigma v_0^y}{ (\beta-\gamma)+\sqrt{(\beta-\gamma)^2+4 \sigma^2}}\right)^T \right\} ,\\[5pt]
\nonumber
L_0^3= & \left\{ \lambda_0^{(3)}=\frac{(\beta+\gamma)-\sqrt{(\beta-\gamma)^2+4 \sigma^2}}{2}, \right. \\
\label{eq:ev3Sol0}
\quad & \hspace{1cm} \left.  w_0=\left( 0,w_0^y,\frac{2 \sigma w_0^y}{ (\beta -\gamma) - \sqrt{(\beta-\gamma)^2+4 \sigma^2}} \right)^T \right\} .
\end{align}
The three results differentiate between the three eigensystems. For simplicity, let us continue with the axion eigenvalue from here. In the next step, we use the result $L_0^1$ to determine $L_1^1$ and $L_2^1$ consecutively:
\begin{align}
\nonumber
\Xi_1^{(1)} &=0, \quad\Delta_1^{(1)} =0,\quad \Theta_1^{(1,2)}=0,\Theta_1^{(1,3)}=0 \\
\Rightarrow\quad L_1^1&=\left\{\lambda_0^{(1)}=0,\lambda_1^{(1)}=0, \vec{u}_0=(1,0,0)^T, 
\vec{u}_1= \left(0, \frac{-\gamma \kappa + \sigma \nu}{\beta \gamma - \sigma^2}, \frac{\kappa\sigma - \beta \nu}{\beta \gamma - \sigma^2 }\right) ^T\right\},\\[10pt]
\nonumber
\Xi_2^{(1)} &=0, \quad\Delta_2^{(1)} =0,\quad \Theta_2^{(1,2)}=0,\Theta_2^{(1,3)}=0 \\
\nonumber
\Rightarrow \quad L_2^1&=\left\{\lambda_0^{(1)}=0,\lambda_1^{(1)}=0,\lambda_2^{(1)} = \alpha - \frac{\gamma \kappa^2 - 2 \sigma \kappa \nu + \beta \nu^2}{\beta \gamma - \sigma^2}, \vec{u}_0=(1,0,0)^T,\right. \\
\nonumber
&\qquad \vec{u}_1= \left(0, \frac{-\gamma \kappa + \sigma \nu}{\beta \gamma - \sigma^2}, \frac{\kappa\sigma - \beta \nu}{\beta \gamma - \sigma^2 }\right) ^T\ ,\\
\label{eq:ev1Sol2}
&\left. \qquad \vec{u}_2=\left(-\frac{\kappa^2(\gamma^2+\sigma^2) - 2 \kappa \sigma \nu ( \beta + \gamma ) + \nu^2 ( \beta^2 +\sigma^2 )}{2(\beta \gamma - \sigma^2 )^2},0,0\right)^T\right\}.
\end{align}
The eigenvalue result for the axion eigensystem is as expected: We have the diagonal term together with a mixing correction at order $1/F_a^2$. We could continue this procedure indefinitely, the only difficulty being the algebraic computations resulting from the constraining equations $\Xi, \Lambda$ and $\Theta$. In a similar way we obtained the second and third eigensystem $L_1^2, L_1^3$ up to order $1/F_a$. 

Some general relations for the eigenvalues $\lambda_0^{(2)}$ and $\lambda_0^{(3)}$, which are especially relevant for the leading pion and $\eta'$ meson masses in the next section, are given by
\begin{align}
    (\lambda_0^{(3)}+\lambda_0^{(2)})^2&=(\beta+\gamma)^2\ ,\\
    (\lambda_0^{(3)}-\lambda_0^{(2)})^2&=(\beta-\gamma)^2+4\sigma^2\ ,\\
    \lambda_0^{(3)}\lambda_0^{(2)}&=\frac{(\lambda_0^{(3)}+\lambda_0^{(2)})^2-(\lambda_0^{(3)}-\lambda_0^{(2)})^2}{4}=\beta\gamma-\sigma^2 \ .
\label{eq:lambdalambda}
\end{align}
With these relations we can reexpress the eigenvalue $\lambda_2^{(1)}$ by
\begin{align}
    \lambda_2^{(1)}F_a^2&=\alpha-\frac{\gamma \kappa^2 - 2 \sigma \kappa \nu + \beta \nu^2}{\beta \gamma - \sigma^2}=\alpha-\frac{\gamma \kappa^2 - 2 \sigma \kappa \nu + \beta \nu^2}{\lambda_0^{(3)}\lambda_0^{(2)}}\label{eq:axexakt}.
\end{align}
\subsection{Applications of the B1 appendix and series expansions of eigenvalues}
\label{subsec:genMatrix3x3Expansions}

Having now established a general expression for the eigenvalues and eigenvectors of a 3$\times$3 matrix expanded in powers of $1 / F_a$, we now apply our expression to the specific mass matrices relevant for the models we consider in the main text.

Our mass matrix follows~\eqnref{massMatrixentries}, where 
\begin{align}
\alpha &= F_a^2 (M^2)_{11} =
\left( v^3 ( m_u (c_2^u)^2 + m_d (c_2^d)^2 ) + (c_3^G)^2 \left( \Lambda_{\eta'}^4 + 2\mu \Lambda_{\text{inst}}^3 \right) \right) \ , \nonumber \\
\kappa &= F_a (M^2)_{12} =
\frac{1}{F_{\eta'}}\left(  m_{c^+} v^3  - c_3^G \left( 2 \Lambda_{\eta'}^4 +  2 \mu \Lambda_{\text{inst}}^3 \right) \right) \ , \nonumber \\
\nu &= F_a (M^2)_{13} = 
\frac{- m_{c^-} v^3 }{F_{\pi^0}} \ , \nonumber \\
\beta &= (M^2)_{22} = 
\frac{1}{F_{\eta'}^2}(m_+ v^3 + 4\Lambda_{\eta'}^4 + 2\mu  \Lambda_{\text{inst}}^3) \ , \nonumber \\
\sigma &= (M^2)_{23} =
\frac{-m_- v^3}{F_{\pi^0} F_{\eta'}} \ , \nonumber \\
\label{eq:translation3x3}
\gamma &= (M^2)_{33} = 
\frac{1}{F_{\pi^0}^2}( m_+ v^3 + 2\mu \Lambda_{\text{inst}}^3) \ .
\end{align}
We identify the $\lambda^{(1)},\lambda^{(2)}$ and $\lambda^{(3)}$ eigenvalues as the axion, $\eta'$ and $\pi^0$ mass eigenvalue, respectively. Putting in the parametrization in \eqnref{translation3x3} into \eqnref{ev2Sol0}, \eqref{eq:ev3Sol0} and \eqref{eq:ev1Sol2} we get
\begin{align}
    m_{\pi^0, \eta'}^2 &= \frac{m_+v^3+4\Lambda_{\eta'}^4+ 2\mu  \Lambda_{\text{inst}}^3}{2F_{\eta'}^2} + \frac{m_+v^3+ 2\mu \Lambda_{\text{inst}}^3}{2F_{\pi^0}^2} \nonumber\\
    \label{eq:pionEtaMass3x3App}
    &\mp \sqrt{\left(\frac{m_+v^3+4\Lambda_{\eta'}^4+ 2\mu  \Lambda_{\text{inst}}^3}{2F_{\eta'}^2} - \frac{m_+ v^3 + 2\mu \Lambda_{\text{inst}}^3}{2F_{\pi^0}^2} \right)^2 + \frac{m_-^2v^6}{F_{\eta'}^2 F_{\pi^0}^2}} \ , \\
    m_a^2F_a^2&=\left( v^3 ( m_u (c_2^u)^2 + m_d (c_2^d)^2 ) + (c_3^G)^2 \left( \Lambda_{\eta'}^4 + 2\mu \Lambda_{\text{inst}}^3 \right) \right)\nonumber\\
    &- \frac{( m_+ v^3 + 2\mu \Lambda_{\text{inst}}^3) \left(  m_{c^+} v^3  - c_3^G \left( 2 \Lambda_{\eta'}^4 +  2 \mu \Lambda_{\text{inst}}^3 \right) \right)^2 }{(m_+ v^3 + 4\Lambda_{\eta'}^4 + 2\mu  \Lambda_{\text{inst}}^3) ( m_+ v^3 + 2\mu \Lambda_{\text{inst}}^3) - (m_- v^3)^2}\nonumber\\
    &+ \frac{  2 m_- v^3 \left(  m_{c^+} v^3  - c_3^G \left( 2 \Lambda_{\eta'}^4 +  2 \mu \Lambda_{\text{inst}}^3 \right) \right)  m_{c^-} v^3 }{(m_+ v^3 + 4\Lambda_{\eta'}^4 + 2\mu  \Lambda_{\text{inst}}^3) ( m_+ v^3 + 2\mu \Lambda_{\text{inst}}^3) - (m_- v^3)^2}\nonumber\\
    &- \frac{ (m_+ v^3 + 4\Lambda_{\eta'}^4 + 2\mu  \Lambda_{\text{inst}}^3) ( m_{c^-} v^3)^2}{(m_+ v^3 + 4\Lambda_{\eta'}^4 + 2\mu  \Lambda_{\text{inst}}^3) ( m_+ v^3 + 2\mu \Lambda_{\text{inst}}^3) - (m_- v^3)^2} \ .\label{eq:axionMass3x3App}
\end{align}
This provides a detailed explanation for~\eqnref{pionEtaMass3x3} and \eqnref{axionMass3x3Extra} in the main text. The axion mass simplifies for the KSVZ model by putting in $c_2^u = 0 =c_2^d$ and $c_3^G=1$ to
\begin{align}
    (m_a^2 F_a^2)^\text{KSVZ} = \Lambda_{\eta'}^4+2\mu  \Lambda_{\text{inst}}^3-\frac{(2\Lambda_{\eta'}^4+2\mu  \Lambda_{\text{inst}}^3)^2(m_+v^3+ 2\mu  \Lambda_{\text{inst}}^3)}{F_{\pi^0}^2m_{\pi^0}^2F_{\eta'}^2m_{\eta'}^2} \ ,
\label{eq:KSVZaxionMassApp}
\end{align}
and for the DFSZ model by putting in $c_3^G=0$ to
\begin{align}
    (m_a^2F_a^2)^\text{DFSZ}&=\frac{4\mu\Lambda_\text{inst}^3(2\Lambda_{\eta'}^4+\mu\Lambda_\text{inst}^3)((c_2^d)^2m_d+(c_2^u)^2m_u)v^3}{F_{\pi^0}^2m_{\pi^0}^2F_{\eta'}^2m_{\eta'}^2}\nonumber\\
    &\quad +\frac{4m_um_dv^6((c_2^d+c_2^u)^2\Lambda_{\eta'}^4+((c_2^d)^2+(c_2^u)^2)\mu\Lambda_\text{inst}^3)}{F_{\pi^0}^2m_{\pi^0}^2F_{\eta'}^2m_{\eta'}^2} \ ,
\label{eq:DFSZaxionMassApp}
\end{align}
which provides the results presented in~\eqnref{KSVZaxionMass} and \eqnref{DFSZaxionMass}. We have now derived all of the needed equations in the main text for Sec.~\ref{subsec:method}. 


In the following, we discuss the KSVZ model in more detail and derive expansions of the result in~\eqnref{KSVZaxionMassApp}. For the KSVZ model we have $c_2^u=0=c_2^d$ and hence $\nu \sim m_{c_-} = 0$.  We expand the eigenvalues of the general mass matrix~\eqref{eq:genMassMatrix} in the isospin conserving limit, where $\sigma \to 0$, as well as the limit $\Lambda_{\eta'} \to \infty$.
The expansions of the mass eigenvalues in the KSVZ model in the isospin limit are given by
\begin{align}
    \lambda_0^{(2)} &= \frac{(\beta + \gamma) + \sqrt{(\beta - \gamma )^2 + 4  \sigma^2}}{2} = \beta - \frac{\beta - \gamma}{2}\left( 1 - \sqrt{ 1 + \frac{ 4 \sigma^2}{( \beta - \gamma)^2}} \right) \nonumber \\
    &= \beta + \frac{ \beta - \gamma}{2} \sum_{k=1}^\infty {1/2 \choose k} \left( \frac{ 4 \sigma^2 }{( \beta - \gamma)^2} \right)^k \label{eq:etapexpansion} \ , \\
    \lambda_0^{(3)} &= \frac{( \beta + \gamma ) - \sqrt{( \beta - \gamma)^2 + 4  \sigma^2}}{2} = \gamma + \frac{ \beta - \gamma}{2} \left( 1 - \sqrt{ 1 + \frac{4 \sigma^2}{( \beta - \gamma)^2}} \right) \nonumber \\
    &= \gamma - \frac{ \beta - \gamma }{2} \sum_{k=1}^\infty {1/2 \choose k} \left( \frac{4 \sigma^2}{ ( \beta - \gamma )^2} \right)^k \label{eq:pi0expansion} \ , \\
    \lambda_2^{(1)} F_a^2 &= \alpha - \frac{\kappa^2}{\beta} \frac{1}{ 1 - \frac{\sigma^2}{\beta \gamma}} = \alpha - \frac{\kappa^2}{\beta} \sum_{k=0}^\infty \left( \frac{\sigma^2}{\beta \gamma} \right)^k = \frac{ \alpha \beta - \kappa^2}{\beta} - \frac{\kappa^2}{\beta} \sum_{k=1}^\infty \left( \frac{\sigma^2}{\beta\gamma} \right)^k \label{eq:axexpansion1} \ ,
\end{align}
where we use the expansion $\sqrt{1+x} = \sum_{k = 0}^{\infty} {1/2 \choose k} x^k$ for the first two results. Using the results in~\eqnsref{etapexpansion}{pi0expansion} and the parametrization in~\eqnref{translation3x3} we obtain the expansions of the $\eta'$ and $\pi^0$ mass eigenvalue used in the main text:
\begin{align}
m_{\eta'}^2 &= \frac{m_+v^3 + 4\Lambda_{\eta'}^4 + 2\mu  \Lambda_{\text{inst}}^3}{F_{\eta'}^2} + \frac{\Delta_m^2}{2} \sum_{k=1}^{\infty} {1/2\choose k} \left( \frac{4m_-^2v^6}{ \Delta_m^4 F_{\pi^0}^2 F_{\eta'}^2} \right)^k 
\label{eq:etaMassexpansionApp} \ , \\
m_{\pi^0}^2 &= \frac{m_+v^3 + 2\mu \Lambda_{\text{inst}}^3}{F_{\pi^0}^2} -\frac{\Delta_m^2}{2} \sum_{k=1}^{\infty} {1/2\choose k} \left( \frac{4 m_-^2 v^6}{\Delta_m^4 F_{\pi^0}^2 F_{\eta'}^2} \right)^k  \label{eq:KimPionMassApp} \ .
\end{align}
For the other limit $\Lambda_{\eta'} \to \infty$ we relate $\Lambda_{\eta'}^4$ to the general parameters in the following way:
\begin{align}
    \Lambda_{\eta'}^4 &= \frac{\beta F_{\eta'}^2 - \gamma F_{\pi_0}^2}{4} = - ( \alpha + \kappa F_{\eta'}) \nonumber \\
    \Leftrightarrow \beta F_{\eta'}^2 &= \gamma F_{\pi_0}^2 - 4(\alpha + \kappa F_{\eta'}) \label{eq:betarel} \ .
\end{align}
Thus, we expand $\lambda_2^{(1)}$ in $(\alpha+\kappa F_{\eta'})$ as is shown in the following calculation:
\begin{align}
    \lambda_2^{(1)} F_a^2 &= \alpha - \frac{ \gamma \kappa^2 }{ \beta \gamma - \sigma^2} = \frac{ \alpha ( \beta \gamma - \sigma^2) F_{\eta'}^2 - \gamma \kappa^2 F_{\eta'}^2}{ ( \beta \gamma - \sigma^2) F_{\eta'}^2} = \frac{ \alpha ( \gamma^2 F_{\pi_0}^2 - \sigma^2 F_{\eta'}^2 ) - \gamma ( 2 \alpha + \kappa F_{\eta'})^2}{( \gamma^2 F_{\pi_0}^2 - \sigma^2 F_{\eta'}^2) - 4 \gamma ( \alpha + \kappa F_{\eta'})} \nonumber  \\
    &= \frac{((2 \alpha + \kappa F_{\eta'}) - ( \alpha + \kappa  F_{\eta'}))( \gamma^2 F_{\pi_0}^2 - \sigma^2 F_{\eta'}^2 ) - \gamma ( 2 \alpha + \kappa F_{\eta'})^2}{- 4 \gamma ( \alpha + \kappa  F_{\eta'})} \frac{1}{ 1 - \frac{ \gamma^2 F_{\pi_0}^2 - \sigma^2 F_{\eta'}^2 }{ 4 \gamma ( \alpha + \kappa F_{\eta'})}} \nonumber \\
    &= \frac{ \gamma^2 F_{\pi_0}^2 - \sigma^2 F_{\eta'}^2}{4 \gamma} + \frac{ \gamma^2 F_{\pi_0}^2 - \sigma^2 F_{\eta'}^2}{4 \gamma} \sum_{k=1}^\infty \left( \frac{ \gamma^2 F_{\pi_0}^2 - \sigma^2 F_{\eta'}^2 }{ 4 \gamma ( \alpha + \kappa F_{\eta'})} \right)^k \nonumber\\
    & \quad + \frac{\gamma( 2 \alpha + \kappa  F_{\eta'})^2 - (\gamma^2 F_{\pi_0}^2 - \sigma^2 F_{\eta'}^2 ) ( 2 \alpha + \kappa F_{\eta'})}{4 \gamma ( \alpha + \kappa  F_{\eta'})} \sum_{k=0}^\infty \left( \frac{ \gamma^2 F_{\pi_0}^2 - \sigma^2 F_{\eta'}^2 }{ 4 \gamma ( \alpha + \kappa  F_{\eta'})} \right)^k \nonumber \\
    &= \frac{ \gamma^2 F_{\pi_0}^2 - \sigma^2 F_{\eta'}^2 }{ 4 \gamma} + \frac{( \gamma^2 F_{\pi_0}^2 - \sigma^2 F_{\eta'}^2 - 2 \gamma( 2 \alpha + \kappa F_{\eta'}))^2}{ 16 \gamma^2 ( \alpha + \kappa F_{\eta'})} \sum_{k=0}^\infty \left( \frac{ \gamma^2 F_{\pi_0}^2 - \sigma^2 F_{\eta'}^2} { 4 \gamma ( \alpha + \kappa F_{\eta'})} \right)^k \label{eq:axexpansion2} \ ,
\end{align}
where we used~\eqnref{betarel} to get rid of the $\beta$ parameter. Using the above result and the parametrization in~\eqnref{translation3x3} we obtain the expansion of the axion mass eigenvalue used in the main text:
\begin{align}
m_a^2 F_a^2 &= \frac{(m_+v^3 + 2\mu \Lambda_{\text{inst}}^3)^2 - m_-^2 v^6}{4(m_+v^3 + 2\mu \Lambda_{\text{inst}}^3)} \nonumber\\
&-\frac{(4\mu^2 \Lambda_{\text{inst}}^6 + m_+^2 v^6 - m_-^2 v^6)^2}{16 \Lambda_{\eta'}^4 (m_+v^3 + 2\mu \Lambda_{\text{inst}}^3)^2} \sum_{k=0}^{\infty} \left( \frac{m_-^2 v^6 - (m_+v^3 + 2\mu \Lambda_{\text{inst}}^3)^2}{ 4\Lambda_{\eta'}^4 (m_+v^3 + 2\mu \Lambda_{\text{inst}}^3)} \right)^k \ . \label{eq:axionExpansionApp}
\end{align}
We have now derived all of the needed equations in the main text for Sec.~\ref{subsec:comparison}.

\subsection{4-dimensional mass matrix}
\label{subsec:genMatrix4x4}
In the following, we parametrize a general $4\times 4$ mass matrix that contains two expansion parameters, where we differ between the two cases with two equal expansion parameters and two different ones. We define the parametrizations to be
\begin{align}
\nonumber
M_{\text{eq}}^{(4\times4)}&=M_0^{(4\times4)}+\frac{1}{F_a}M_1^{(4\times4)}+\frac{1}{F_a^2}M_2^{(4\times4)}\\
&=\left(
\begin{array}{cccc}
0&0&0&0\\
0&  0& 0& 0 \\
 0 &0& \beta & \sigma \\
 0 &0 &\sigma & \gamma \\
\end{array}
\right)+\frac{1}{F_a}\left(
\begin{array}{cccc}
0&0&0&0\\
0& 0 & \kappa & 0 \\
0& \kappa & 0 & 0 \\
 0& 0 & 0 & 0 \\
\end{array}
\right)+\frac{1}{F_a^2}\left(
\begin{array}{cccc}
\alpha_1 &\kappa_1 & 0 & 0 \\
 \kappa_1 &\alpha & 0 & 0 \\
0& 0 & 0 & 0 \\
0&0&0&0\\
\end{array}
\right)
\label{eq:genMassMatrix4x4eqScale}
=\left(
\begin{array}{cccc}
\frac{\alpha_1}{F_a^2} &\frac{\kappa_1}{F_a^2}& 0 & 0 \\
\frac{\kappa_1}{F_a^2} &\frac{\alpha}{F_a^2}  & \frac{\kappa}{F_a} & 0 \\
0& \frac{\kappa}{F_a}& \beta & \sigma \\
 0 &0 &\sigma & \gamma \\
\end{array}
\right),\\[10pt]
\nonumber
M_{\text{diff}}^{(4\times4)}&=M_0^{(4\times4)}+\frac{1}{F_d}M_{1d}^{(4\times4)}+\frac{1}{F_d^2}M_{2d}^{(4\times4)}+\frac{1}{F_aF_d}M_{2ad}^{(4\times4)}+\frac{1}{F_a^2}M_{2a}^{(4\times4)}\\
\nonumber
&=\left(
\begin{array}{cccc}
0&0&0&0\\
0&  0& 0& 0 \\
 0 &0& \beta & \sigma \\
 0 &0 &\sigma & \gamma \\
\end{array}
\right)+\frac{1}{F_d}\left(
\begin{array}{cccc}
0&0&0&0\\
0& 0 & \kappa & 0 \\
0& \kappa & 0 & 0 \\
 0& 0 & 0 & 0 \\
\end{array}
\right)+\frac{1}{F_d^2}\left(
\begin{array}{cccc}
0 &0 & 0 & 0 \\
0 &\alpha & 0 & 0 \\
0& 0 & 0 & 0 \\
0&0&0&0\\
\end{array}
\right)\\
\label{eq:genMassMatrix4x4diffScale}
&\quad +\frac{1}{F_aF_d}\left(
\begin{array}{cccc}
0 &\kappa_1 & 0 & 0 \\
\kappa_1 &0 & 0 & 0 \\
0& 0 & 0 & 0 \\
0&0&0&0\\
\end{array}
\right)
+\frac{1}{F_a^2}\left(
\begin{array}{cccc}
\alpha_1 &0 & 0 & 0 \\
0 &0 & 0 & 0 \\
0& 0 & 0 & 0 \\
0&0&0&0\\
\end{array}
\right)
=\left(
\begin{array}{cccc}
\frac{\alpha_1}{F_a^2} &\frac{\kappa_1}{F_aF_d}& 0 & 0 \\
\frac{\kappa_1}{F_aF_d} &\frac{\alpha}{F_d^2}  & \frac{\kappa}{F_d} & 0 \\
0& \frac{\kappa}{F_d}& \beta & \sigma \\
 0 &0 &\sigma & \gamma \\
\end{array}
\right).
\end{align}
The eigensystem of the matrix $M_{\text{eq}}^{(4\times4)}$ shows a degeneracy between the two axion eigenvalues that is resolved at second order. Omitting the long expressions for the eigenvectors, we present the resulting eigenvalues $M_{\text{eq}}^{(4\times4)}$ of the two axion masses up to second order for $\kappa_1\neq 0$ : 
\begin{align}
\lambda^{(1)}_0=0,\qquad &\lambda^{(1)}_1=0,\nonumber \\
\lambda^{(1)}_2 = & \frac{1}{2}\left( \alpha_1+ \left( \alpha - \frac{\gamma \kappa^2}{\beta \gamma - \sigma^2}\right) + \sqrt{\left( \alpha_1- \left( \alpha - \frac{\gamma \kappa^2}{\beta \gamma - \sigma^2}\right) \right)^2 + 4\kappa_1^2} \right)\ , \nonumber\\
\lambda^{(2)}_0=0,\qquad &\lambda^{(2)}_1=0,\nonumber\\
\lambda^{(2)}_2=\alpha_1 - & 2 \kappa_1^2  \label{eq:ev1Sol24x4} \\
& \times \abs{  \alpha_1 - \left(  \alpha - \frac{\gamma \kappa^2}{\beta \gamma - \sigma^2} \right) - \sqrt{\left(  \alpha_1 -\left(  \alpha - \frac{\gamma \kappa^2}{\beta \gamma - \sigma^2} \right)\right)^2+4\kappa_1^2} }^{-1} \ , \label{eq:ev2Sol24x4} 
\end{align}
where $\lambda^{(n)} = \sum_{i=0}^{\infty} \frac{1}{F_a^i} \lambda_i^{(n)}$ corresponds to the expansion of the first and second axion mass for $n=1$ and $n=2$, respectively. In the case of $\kappa_1 = 0$ we have a trivial eigenvalue and the 3-dim. eigensystem in \eqnref{genMassMatrix}, which results in $\lambda^{(1)}_2=\alpha_1 \ , \lambda^{(2)}_2 = \alpha - \gamma\kappa^2/(\beta\gamma - \sigma^2)$ as in \eqnref{ev1Sol2}. For the other two eigenvalues  $\lambda^{(3)}_0$ and $\lambda^{(4)}_0$ correspond to $\lambda^{(2)}_0$ and $\lambda^{(3)}_0$ in \eqnsref{ev2Sol0}{ev3Sol0}, respectively. The eigenvector solutions to first order in $F_a$ are
\begin{align}
    \vec{u}_0 &= (1,0,0,0)^T \ , \nonumber \\
    \vec{u}_1 &= \frac{1}{\sqrt{\kappa_1^2 + \left(\alpha_1 - \lambda_2^{(1)} \right)^2}} \left( F_a |\alpha_1 - \lambda_2^{(1)}|/\kappa_1 \ , -F_a \ , \frac{\gamma \kappa}{\beta \gamma - \sigma^2} \ , - \frac{\kappa_1 \kappa\sigma}{\beta \gamma - \sigma^2} \right)^T \ ,\label{eq:eVec14x4M1} \\
    \vec{v}_0 &= (0,1,0,0)^T \ , \nonumber \\
    \vec{v}_1 &= \frac{1}{\sqrt{\kappa_1^2 + \left(\lambda_2^{(1)} - \left(\alpha - \frac{\gamma\kappa^2}{\beta \gamma - \sigma^2} \right) \right)^2}} \nonumber \\
    & \qquad \cdot \left( F_a \left( \lambda_2^{(1)} - (\alpha - \frac{\gamma\kappa^2}{\beta \gamma - \sigma^2})\right)/\kappa_1 \ , F_a \ , - \frac{\gamma \kappa}{\beta \gamma - \sigma^2} \ ,  \frac{\kappa_1 \kappa \sigma}{\beta \gamma - \sigma^2} \right)^T \ , \label{eq:eVec24x4M1}
\end{align}
where $\vec{u}=\sum_{i=0}^\infty \frac{1}{F_a}\vec{u_i}$ and $\vec{v}=\sum_{i=0}^\infty \frac{1}{F_a}\vec{v_i}$ are the total eigenvectors of the first and second eigenvalue, respectively.

The eigensystem of the matrix $M_{\text{diff}}^{(4\times4)}$ has no degeneracy, but there is a large number of unknowns in the case with no hierarchy between $F_a$ and $F_d$. For our case, we can set $F_a\gg F_d$, which simplifies the eigensystem significantly. Using the same approach we obtain the resulting eigenvalues for the to second order:
\begin{align}
\lambda^{(n)} &= \sum_{i=0}^{\infty} \left( \frac{1}{F_a^i} (\lambda_a)_{i}^{(n)}  \frac{1}{F_d^i} (\lambda_d)_{i}^{(n)} +  \right) + \sum_{i=1}^{\infty} \sum_{j=1}^{\infty} \left( \frac{1}{F_a^i F_d^j} (\lambda_{ad})_{ij}^{(n)} \right) \\
(\lambda_a)_{0}^{(1)} &= 0 =(\lambda_d)_{0}^{(1)} \ ,\quad (\lambda_a)_{1}^{(1)} = 0 = (\lambda_d)_{1}^{(1)}\ , \quad (\lambda_d)_2^{(1)} = 0 \ , \nonumber\\
(\lambda_a)_2^{(1)} &= \alpha_1 - \frac{\kappa_1^2 }{ \alpha - \frac{\gamma \kappa^2}{\beta \gamma - \sigma^2}} \ , \label{eq:ev1sol24x4mix}\\
(\lambda_a)_{0}^{(2)} &= 0 =(\lambda_d)_{0}^{(2)} \ ,\quad (\lambda_a)_{1}^{(2)} = 0 = (\lambda_d)_{1}^{(2)}\ , \nonumber\\
(\lambda_d)_{2}^{(2)} &= \alpha - \frac{\gamma \kappa^2}{\beta \gamma - \sigma^2} \ , \quad (\lambda_{ad})_{11}^{(2)} = 0 \ , \quad (\lambda_a)_{2}^{(2)} = \frac{\kappa_1^2 }{ \alpha - \frac{\gamma \kappa^2}{\beta \gamma - \sigma^2}}  \label{eq:ev2sol24x4mix}\ ,
\end{align}
where the other two eigenvalues  $\lambda^{(3)}_0, \lambda^{(4)}_0$ correspond to $\lambda^{(2)}_0$ and $\lambda^{(3)}_0$ in \eqnsref{ev2Sol0}{ev3Sol0}, respectively.

\subsection{Applications of the B4 appendix and series expansions of eigenvalues}
\label{subsec:genMatrix4x4Expansions}
In this section we are going to apply the results derived in the above section to $4\times 4$ mass matrices discussed in the main text. 

Our first mass matrix follows~\eqnref{massMatrixGaillard}, where 
\begin{align}
\alpha_1 & = F_a^2 (M_{\text{M1}}^2)_{11} =  4 ( \Lambda_{\text{SSI}}^4 + \Lambda_{\text{diag}}^4)\ , \nonumber \\
\kappa_1 & = F_a^2 (M_{\text{M1}}^2)_{12} =  2 \sqrt{6} \Lambda_{\text{diag}}^4 \ , \nonumber \\
\alpha &= F_a^2 (M_{\text{M1}}^2)_{22} =  6 \left(\Lambda_{\text{diag}}^4 + \Lambda_{\eta'}^4 + 2 \mu \Lambda_{\text{inst}}^3 \right) \ , \nonumber \\
\kappa &= F_a (M_{\text{M1}}^2)_{23} = 
\frac{-2 \sqrt{6}}{F_{\eta'}} ( \Lambda_{\eta'}^4 + \mu \Lambda_{\text{inst}}^3 ) \ , \nonumber \\
\beta & = (M_{\text{M1}}^2)_{33} =  
\frac{1}{F_{\eta'}^2}(m_+ v^3 + 4\Lambda_{\eta'}^4 + 2\mu  \Lambda_{\text{inst}}^3) \ , \nonumber \\
\sigma &= (M_{\text{M1}}^2)_{34} = 
\frac{-m_- v^3}{F_{\pi^0} F_{\eta'}} \ , \nonumber \\
\label{eq:translation4x4}
\gamma &= (M_{\text{M1}}^2)_{44} =  
\frac{1}{F_{\pi^0}^2}( m_+ v^3 + 2\mu \Lambda_{\text{inst}}^3) \ .
\end{align}
The new mass matrix includes the previous one in~\eqnref{translation3x3} with an additional factor $6$ for the axion eigenvalue. Using the parametrization from~\eqnref{translation4x4} this can be expressed as
\begin{align}
     6(m_a^2 F_a^2)^\text{KSVZ} = \alpha - \frac{\gamma \kappa^2}{\beta\gamma - \sigma^2} - 6 \Lambda_{\text{diag}}^4 \ , \label{eq:simplificationKSVZaxion}
\end{align}
where $(m_a^2 F_a^2)^\text{KSVZ}$ corresponds to the solution presented in~\eqnref{KSVZaxionMassApp}.
We identify the $\lambda^{(1)},\lambda^{(2)},\lambda^{(3)}$ and $\lambda^{(4)}$ eigenvalues as the axion, axieta, $\eta'$ and $\pi^0$ mass eigenvalue, respectively. Note that the assignment of $\lambda^{(1)},\lambda^{(2)}$ can interchange depending on the explicit values for $\Lambda_{\text{SSI}}$ and $ \Lambda_{\text{diag}}$. Putting in the parametrization in \eqnref{translation4x4} into \eqnsref{ev1Sol24x4}{ev2Sol24x4} we get
\begin{align}
    m_{\eta_d}^2F_a^2 =  & 4 \Lambda_{\text{SSI}}^4 + 10 \Lambda_{\text{diag}}^4 + 6(m_a^2 F_a^2)^\text{KSVZ} \nonumber\\
    & \qquad  + \sqrt{ \left(4 \Lambda _{\text{SSI}}^4 - 2 \Lambda_{\text{diag}}^4 - 6(m_a^2 F_a^2)^\text{KSVZ}\right)^2 + 96 \Lambda_{\text{diag}}^8 }  \label{eq:GaillardaxionmassApp} \ , \\
    m_{a}^2F_a^2 = &  4( \Lambda_{\text{diag}}^4+\Lambda _{\text{SSI}}^4) - 48 \Lambda_{\text{diag}}^8 \nonumber \\
    & \cdot \abs{ 4 \Lambda_{\text{SSI}}^4 - 2 \Lambda_{\text{diag}}^4 - 6(m_a^2 F_a^2)^\text{KSVZ}  - \sqrt{ \left(4 \Lambda _{\text{SSI}}^4 - 2 \Lambda_{\text{diag}}^4 - 6(m_a^2 F_a^2)^\text{KSVZ}\right)^2 + 96 \Lambda_{\text{diag}}^8 } }^{-1} \label{eq:GaillardEtaDmassApp} \ .
\end{align}
These relations correspond to \eqnsref{Gaillardaxionmass}{GaillardEtaDmass} discussed in the main text in \subsecref{dynAxCol}. 

The second mass matrix has two different expansion parameters and follows ~\eqnref{massMatrix2GaillardEntries}, where
\begin{align}
\alpha_1 & = F_a^2 (M_{\text{M2}}^2)_{11} =  144 ( \Lambda_{\text{SSI}}^4 + \Lambda_{\text{diag}}^4)\ , \nonumber \\
\kappa_1 & = F_d^2 (M_{\text{M2}}^2)_{12} =  2 \sqrt{6} \Lambda_{\text{diag}}^4 \ , \label{eq:translation4x4mix}
\end{align}
and the rest is the same as in~\eqnref{translation4x4}. 
Applying ~\eqnsref{ev1sol24x4mix}{ev2sol24x4mix} to~\eqnref{translation4x4mix} and using~\eqnref{simplificationKSVZaxion} gives
\begin{align}
    m_{a'}^2 F_a^2 = & 144 \left(\Lambda_{\text{diag}}^4 + \Lambda _{\text{SSI}}^4 - \Lambda_{\text{diag}}^4 \left(1 + \frac{(m_a^2 F_a^2)^\text{KSVZ}}{\Lambda_{\text{diag}}^4}\right)^{-1} \right)  \ , \label{eq:Gaillard2AxionPrimeMassApp}\\
    m_{a}^2 F_d^2 = & 6 \Lambda_{\text{diag}}^4 + 6 (m_a^2 F_a^2)^\text{KSVZ} + \frac{F_d^2}{F_a^2} 144 \left( \Lambda_{\text{diag}}^4 \left( 1 + \frac{(m_a^2 F_a^2)^\text{KSVZ}}{\Lambda_{\text{diag}}^4} \right)^{-1} \right)\ . \label{eq:Gaillard2AxionMassApp}
\end{align}